\documentclass[twocolumn]{aastex631}
\usepackage{hyperref, amsmath, latexsym, graphicx, amssymb, longtable, epsf, enumitem, float, natbib}
\hypersetup{linkcolor=blue, citecolor=blue, filecolor=magenta, urlcolor=blue}

\newcommand{\planck}{\textit{Planck}\ }

\graphicspath{{./}{figures/}}

\newcommand{\lcdm}{$\Lambda$CDM}

\begin{document}
\title{Dwarf Galaxy Number Counts within 25 Mpc: Predictions from Local Group Analogues in TNG50} 

\correspondingauthor{Evangela Shread}
\email{eshread@caltech.edu}

\author[0000-0003-1211-530X]{Evangela~E.~Shread}
\affiliation{Cahill Center for Astronomy \& Astrophysics, California Institute of Technology, Pasadena, CA 91125, USA}

\author[0000-0003-1661-2338]{Joanna~M.~Piotrowska}
\affiliation{Cahill Center for Astronomy \& Astrophysics, California Institute of Technology, Pasadena, CA 91125, USA}

\author[0000-0002-6442-6030]{Daniel~R.~Weisz}\affil{Department of Astronomy, University of California, Berkeley, CA 94720, USA}

\author[0000-0002-9604-343X]{Michael Boylan-Kolchin}
\affiliation{Department of Astronomy, The University of Texas at Austin, Austin, TX 78712, USA}
\affiliation{Cosmic Frontier Center, The University of Texas at Austin, Austin, TX 78712, USA}
\affiliation{Weinberg Institute for Theoretical Physics, The University of Texas at Austin, Austin, TX 78712, USA}

\author[0000-0002-4226-8959]{Fiona~A.~Harrison}\affiliation{Cahill Center for Astronomy \& Astrophysics, California Institute of Technology, Pasadena, CA 91125, USA}

\begin{abstract} 
    The modern generation of wide-field galaxy surveys, such as LSST, Euclid and Roman, will enable studies of dwarf galaxies ($10^6\leq M_\ast/M_\odot \leq10^9$) beyond the Local Group (LG) in unprecedented detail. Improved theoretical understanding of this population is necessary to guide these observations, since predictions in this regime are generally limited to specific environments like the LG. We present predictions for the population of dwarf galaxies from the TNG50 run of the IllustrisTNG suite of cosmological hydrodynamical simulations, focusing on the environments within $1<D/\mathrm{Mpc}<25$ of LG analogues at $z=0$. In the simulated sample, there are $\sim1,000$ and $\sim12,000$ dwarf galaxies within $10$ and $25~\mathrm{Mpc}$, respectively. We compare our results with the 50 Mpc Galaxy Catalog and estimate that current observations are highly incomplete at low masses: for $10^6 \, \leq \, M_\ast/M_\odot \, \leq \, 10^7$ ($-13\lesssim M_r\lesssim-10$), we find completeness fractions of $\sim23\%$ within $10$ Mpc and $\sim4\%$ within $25$ Mpc. The simulated galaxies below the completeness limits of the observations exist in a range of environments, with notable populations of field dwarfs at all distances and satellites around centrals with masses $10^8\lesssim M_\ast/M_\odot \lesssim10^{11}$ within $10-25$ Mpc. We find that there are $\sim8$ times more quiescent dwarf galaxies in the TNG50 sample than are currently cataloged. Our results suggest that upcoming observations should uncover a substantial population of dwarf galaxies, and that $\gtrsim15\%$ 
    of these will be red, currently quenched galaxies in the field.
\end{abstract} 


\section{Introduction} \label{sec:1}

Low-mass galaxies ($M_\ast \leq 10^9~M_\odot$) are unique laboratories for understanding galaxy formation and evolution from the epoch of reionization to the present, for constraining the physics of stellar evolution and feedback in metal-poor environments (e.g. \citealt{FrebelNorris_2015}), and for testing $\Lambda$ + cold dark matter (\lcdm) cosmology on small scales (e.g. \citealt{LCDMChallengesReview}). They are the most abundant galaxies in the Universe at all redshifts, but our most complete census of dwarf galaxies is limited to the Local Group (LG; D $\lesssim 1 \ \mathrm{Mpc}$), where there has been a particular emphasis on finding and characterizing the faintest galaxies (e.g. \citealt{Willman_2005, McConnachie2012, Bechtol_2015, Drlica-Wagner_2015, Laevens_2015, FaintDwarfsReview2019}). A more complete census of dwarf galaxies at greater distances is crucial to advance our knowledge of these systems.

Substantial progress has been made in recent years, aided by advances in instrumentation and increasingly sophisticated search algorithms. Beyond the LG, new dwarf galaxies within $\sim 10~\mathrm{Mpc}$ can be identified from ground-based imaging using their resolved stellar populations (e.g. \citealt{Carlin_2016, Carlin_2024, Bennet_2017, Crnojevic_2019, Mutlu-Pakdil_2022, dolivadolinsky2025}), while at larger distances most discoveries are made via integrated light, as individual stars are too faint and crowded even for the most powerful wide-area, ground-based telescopes (e.g. \citealt{Muller2017_M101, Danieli_2018, Karachentsev_2022_DESI, Hunter_2025}). Spectroscopic surveys have also enabled the characterization of low-mass galaxies (e.g. \citealt{SAGA2017, SAGA_2021, Darragh-Ford_2023}). 

%Although such deep, targeted studies have uncovered the faintest nearby galaxies and have been critical for advancing our knowledge of low-mass galaxy formation and evolution, they typically lack the sky coverage and sample statistics needed for a census of local dwarf galaxies. 

The discovery space for dwarf galaxies is expected to grow tremendously with ongoing and upcoming wide-field imaging surveys as the Euclid mission \citep{Euclid_Overview}, the Nancy Grace Roman Space Telescope (Roman; \citealt{Roman2015}), and the Vera C. Rubin Observatory's Legacy Survey of Space and Time (LSST; \citealt{LSST2019}). LSST, in particular, will enable resolved and semi-resolved searches throughout the Local Volume (LV; $D\sim 10~\mathrm{Mpc}$) and beyond. The ten-year main survey for LSST is expected to have a footprint of $\sim18,000~\mathrm{deg}^2$, a co-added imaging depth of $r < 27.5$, and an estimated surface brightness limit of $\sim30~\mathrm{mag/arcsec^2}$, representing an unprecedented combination of wide and deep imaging \citep{LSST2019, LSSTsoftware}.

With these efforts underway, an improved theoretical understanding of the expected local dwarf galaxy population is necessary to guide observing strategies and to provide a benchmark against which future observations may be compared. Previous theoretical efforts to constrain the abundance of local low-mass galaxies generally focus on satellites of the Local Group (e.g. \citealt{Sawala2016, GarrisonKimmel2019, Nadler2020}), and so they offer limited predictive power for different environments. Other theoretical predictions are derived from models of the galaxy-halo connection, which are often calibrated at higher masses (e.g. \citealt{Moster2013, Behroozi_2013, Behroozi_2019}) and remain poorly constrained at low masses (e.g. \citealt{Brook_2014, Garrison-Kimmel2014, Allen_2019}). The challenge of estimating the number count and properties of the undiscovered dwarf galaxy population beyond the LG can be addressed with modern cosmological hydrodynamical simulations, which now have the resolution needed for theoretical studies of statistical samples of dwarf galaxies down to masses of $M_\ast\sim 10^6-10^7 \, M_\odot$ in volumes tens of Mpc wide. In these simulations, dark matter and baryonic components are evolved from initial conditions set by the \lcdm\ model, and processes such as star formation and chemical enrichment, active galactic nucleus (AGN) and stellar feedback, and black hole growth and accretion are modeled over a wide range of spatial and temporal scales (see e.g. \citealt{Vogelsberger2020} for a  review). These simulations offer a resource for not only interpreting observations, but also for making predictions for the observable properties of the local galaxy population.  

In this paper, we deliver such predictions for the still-undiscovered population of dwarf galaxies within a radius of $25$ Mpc from the Milky Way (MW). To do so, we explore the distribution and global properties of simulated dwarf galaxy analogs at $z=0$ in the TNG50 run of the IllustrisTNG simulation suite \citep{Nelson_2019_TNG50results, Pillepich_2019}. We focus our study on the three Local Group analogues that were identified by \citet{Pillepich_2024} within the $z=0$ snapshot of the highest resolution run of TNG50 (TNG50-1). Using these simulated environments centered on MW-like galaxies, we can directly compare theoretical predictions for the abundance of local dwarf galaxies with existing observations. We extend our study to a mass limit of $M_\ast \approx 10^{6} \ M_\odot$ to target the classical dwarf population, which the TNG50 simulation box size allows us to study within a volume of $D < 25~\mathrm{Mpc}$.

The aim of this article is to provide theoretical predictions for the number of classical and bright dwarf galaxies ($10^6 \leq M_\ast/M_\odot \leq 10^9$) within $1 \leq D/\mathrm{Mpc} \leq 25$, along with their distributions by group versus field and star-forming versus quenched classifications, based on the state-of-the-art IllustrisTNG simulations. In \S \ref{sec:2}, we describe our selected samples of simulated galaxies and outline our methods for consistently comparing simulation data and observations. In \S \ref{sec:3}, we present our results, and in \S \ref{sec:discussion}, we assess the completeness of current observations and discuss the implications of our work for upcoming surveys, with a focus on LSST. We summarize our main findings in \S \ref{sec:conclusion}. In this work, we adopt the \planck 2016 cosmology\footnote{The \planck 2016 parameters are $h = 0.6774$, $\Omega_{\Lambda,0} = 0.6911$, $\Omega_{m,0} = 0.3089$, $\Omega_{b,0} = 0.0486$, $\sigma_8 = 0.8159$ and $n_s = 0.9667$.} \citep{Planck2016} and assume a Chabrier initial mass function (IMF; \citealt{Chabrier2003}), consistent with the IllustrisTNG model choices.

\begin{deluxetable*}{cccccc}\label{tab:TNG50_MW_properties}
\tablecolumns{6}
\tablecaption{Basic properties of the three MW and M31 simulated galaxy analogue pairs that define the MW/LG analogue samples in this work. These were identified by \citet{Pillepich_2024} in the $z=0$ TNG50 snapshot. Also listed are the number of bright galaxies $N(> 10^{9.5}~M_\odot)$  and cold dark matter density within $1-10~\mathrm{Mpc}$, as defined in \S \ref{results:counts}.}
\tablehead{
\colhead{} & \colhead{} & \colhead{Stellar mass} & \colhead{FoF halo mass} & \colhead{$N(> 10^{9.5} M_\odot)$} & \colhead{$\Omega_\mathrm{cdm,eff}$} \\
\colhead{} & \colhead{TNG50 Subhalo ID} & \colhead{$(M_\ast/M_\odot)$} & \colhead{$(M_{200\mathrm{c}}/M_\odot)$} & \colhead{$(1 < D/\mathrm{Mpc} < 10)$} & \colhead{$(1 < D/\mathrm{Mpc} < 10)$}
}
\startdata 
        MW A & 400974 & $3.05 \times 10^{10}$ & $3.77 \times 10^{12}$ & 48 & 0.29 \\
        M31 A & 400973 & $1.09 \times 10^{11}$ & $3.77 \times 10^{12}$ & & \\ \hline        
        MW B & 454172 & $3.60 \times 10^{10}$ & $1.63 \times 10^{12}$ & 71 & 0.40 \\
        M31 B & 454171 & $5.90 \times 10^{10}$ & $1.63 \times 10^{12}$ & & \\ \hline
        MW C & 580406 & $2.59 \times 10^{10}$ & $4.99 \times 10^{11}$ & 103 & 0.61 \\
        M31 C & 425719 & $8.61 \times 10^{10}$ & $3.33 \times 10^{12}$ & & 
    \enddata
\end{deluxetable*}

\section{Sample Selection and Data}
\label{sec:2}

\subsection{IllustrisTNG Cosmological Simulation Suite}

The TNG50 simulation bridges the gap between large-volume and zoom-in simulations, as it is the highest-resolution run of the IllustrisTNG suite of \lcdm\  magnetohydrodynamical (MHD) cosmological simulations \citep{Marinacci_2018, Naiman_2018, Nelson_2017, Pillepich_2017_TNGintrostellarcontent, Springel_2017}. With a box size of $35 \ h^{-1} \approx 50$ comoving Mpc on each side and a mass resolution of $8.5 \times 10^4 \, M_\odot$ for baryonic matter and $4.5 \times 10^5 \, M_\odot$ for dark matter, TNG50 offers a combination of volume and mass resolution that is well suited for studies of the properties of the nearby dwarf galaxy population. The TNG baryonic physics model includes sub-resolution (subgrid) prescriptions for gas cooling and heating, chemical enrichment, star formation, supernova feedback, AGN feedback, and black hole growth (see e.g. \citealt{Pillepich_2017_TNGintrogalaxyformation, Weinberger_2016}). Across the entire suite, it has been shown to reproduce fundamental properties and scaling relations of observed galaxies, including, e.g., the shape and normalization of the stellar mass functions (SMFs) at low and intermediate redshifts as well as the shapes and widths of the red and blue sequences of Sloan Digital Sky Survey (SDSS) galaxies \citep{Nelson_2017, Pillepich_2017_TNGintrostellarcontent, Springel_2017, Marinacci_2018, Naiman_2018}. 

The subgrid models in the IllustrisTNG suite are calibrated to reproduce select observational results, such as the star formation rate density as a function of cosmic time, the galaxy SMF at $z \sim 0$, and the stellar-to-halo mass (SHMR) relation at $z = 0$ \citep{Pillepich_2017_TNGintrogalaxyformation}. The observational SMF results used to calibrate the TNG model only extend down to $M_\ast \sim 10^{8} \, M_\odot$, so the SMF at lower masses can be considered a prediction of the TNG models into an extrapolated regime. 

The simulated galaxies in TNG50 belong to subhalos which are identified using the SUBFIND algorithm developed by \citet{SUBFIND}. This routine extracts locally overdense, self-bound dark matter substructures from larger, virialized halos in simulations. It is based on the friends-of-friends algorithm (FoF; \citealt{FoF1985}) which identifies groups of particles according to a specified linking length multiplied by the mean interparticle distance. All subhalos belong to a particular FoF group, which we refer to as the host halo. Note that throughout this paper, we use the terms `subhalo' and `galaxy' interchangeably, as the galaxy occupation fraction for subhalos in the mass range we are considering is expected to be 100\% (e.g. \citealt{BenitezLlambay2020}). We apply our own method of classifying galaxy groups, which we discuss in Appendix \ref{sec:appendixGF}.

\begin{figure*}[ht]
    \centering
    \includegraphics[width=\linewidth]{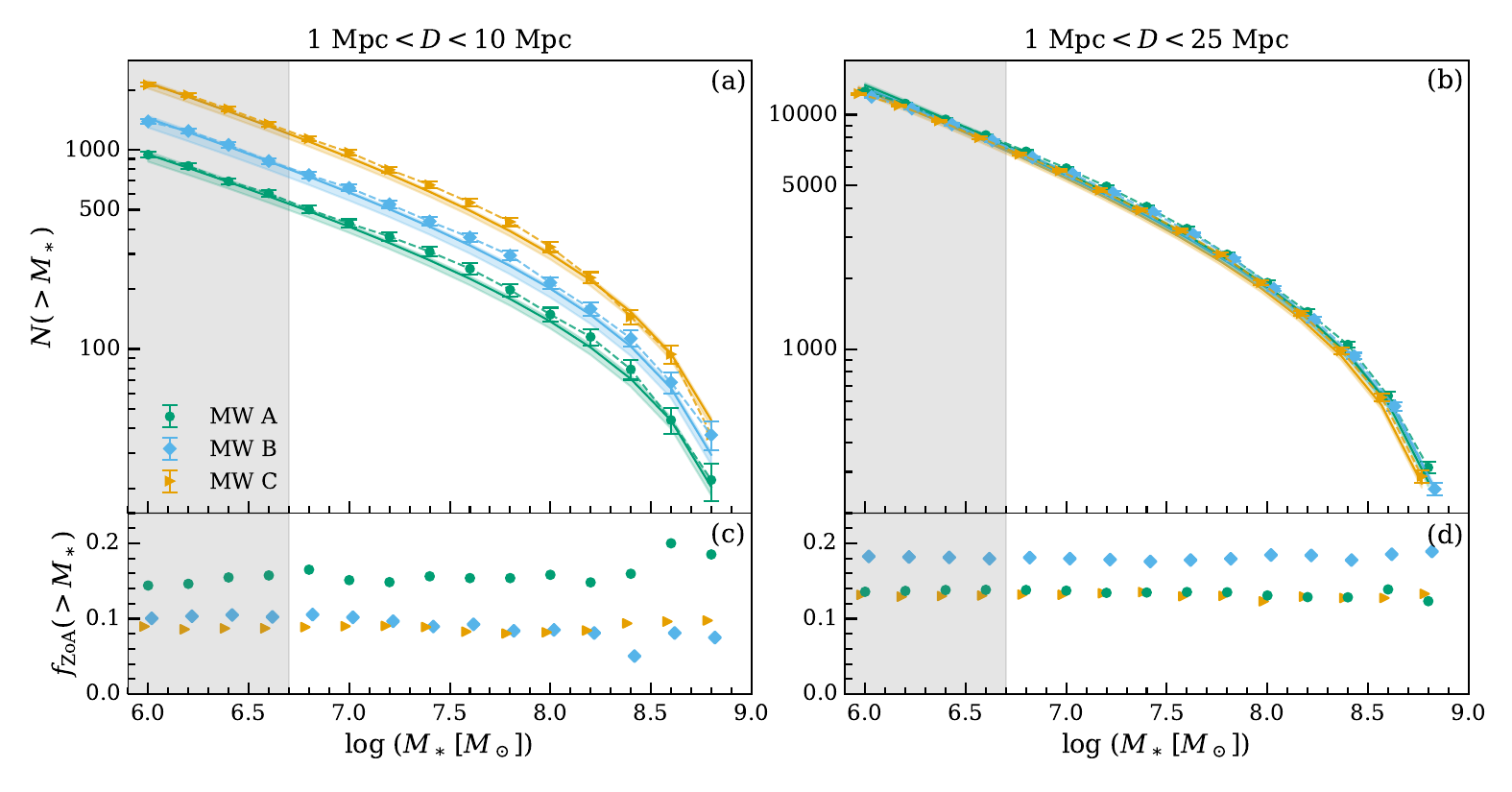}
    \caption{(\textit{Top row}) Cumulative number counts as a function of mass for the three TNG50 MW analogue samples, within (a) $1 < D/\mathrm{Mpc} < 10$ and (b) $1 < D/\mathrm{Mpc} < 25$. Markers and dashed lines show the raw counts, with error bars representing the Poisson counting error. Counts are computed in bins spaced $\Delta\log(M_\ast/M_\odot) = 0.2 \ \mathrm{dex}$ apart, and only galaxies with masses $10^6 \leq M_\ast/M_\odot \leq 10^9$ are counted. Solid lines show the integrated Schechter fits to the mass functions in each volume, with shaded errors derived from the fit parameter uncertainties. (\textit{Bottom row}) A `Zone of Avoidance' (ZoA) in the region $|b| \leq 10^\circ$ is defined relative to the direction of net stellar angular momentum of each MW analogue, and all subhalos in this region are removed from the surrounding volume. The cumulative fraction of galaxies removed as a function of mass $f_\mathrm{ZoA}(> M_\ast)$ is shown within (c) $1 < D/\mathrm{Mpc} < 10$ and (d) $1 < D/\mathrm{Mpc} < 25$. In all plots, the gray region indicates where there are fewer than 100 stellar particles in the TNG50 subhalos. In (b) and (d), a small offset in $\log M_\ast$ is applied to the MW B and C data relative to MW A for easier visualization.}\label{fig:TNG-counts}
\end{figure*}

\subsubsection{MW and M31 Analogues in TNG50}\label{2.1}
    
\citet{Pillepich_2024} identified 198 individual MW or M31 analogues in the $z=0$ snapshot of TNG50. These analogue galaxies are selected based on the following criteria: 
\begin{enumerate}[label=(\arabic*)]
    \item The MW or M31 analogue has a disky stellar morphology and a stellar mass of $10.5 \leq \log (M_\ast/M_\odot) \leq 11.2$ contained within a radius of $30 \ \mathrm{kpc}$;
    \item The total mass of the host halo is $M_{200\mathrm{c}} < 10^{13}\,M_\odot$; and
\item There are no other galaxies with a stellar mass $M_\ast \geq 10^{10.5}\,M_\odot$ within a distance of 500 kpc.
\end{enumerate}
The abundances and properties of the satellites of these simulated galaxies agree reasonably well with observations of the MW and M31 satellite systems. The simulated satellites with masses $M_\ast > 5 \times 10^6 M_\odot$ follow basic observed scaling relations for MW and M31 satellites -- such as stellar velocity dispersion as a function of stellar mass -- and they are consistent with the observed MW and M31 satellite luminosity functions, validating the objects selected as MW/M31 analogues \citep{Engler_2021}. 

Within this sample, \citet{Pillepich_2024} identified three Local Group-like analogues each consisting of a pair of MW- and M31-like galaxies. For any two individual MW or M31 analogues to be considered an LG-like pair, they must meet the following criteria:
\begin{enumerate}[label=(\arabic*)]
    \item The two analogue galaxies must be within $500-1000 \ \mathrm{kpc}$ of one another, and there cannot be any other MW or M31 analogues within this distance; and
    \item The two galaxies must have a negative radial velocity with respect to one another.
\end{enumerate}

Most importantly for our study, the TNG simulations are not tuned to produce LG-like pairs, and so studying the environments around these systems constitutes a prediction for the very local galaxy population. There is theoretical motivation to believe that the three LG-like systems in TNG50 live in large-scale environments similar to the Local Volume. Previous studies of LG analogues identified in other cosmological simulations, which also do not impose additional constraints to enforce the formation of LG-like systems, suggest that these systems emerge in similar large-scale environments (e.g. \citealt{Forero-Romero_2015, Zhai2020}). This conclusion is supported by simulations tuned to reproduce the properties of the local Universe (e.g.  \citealt{Sorce2016, Libeskind2020}). Notably, \citet{Carlesi2016} contend that the formation of the LG is a robust outcome of \lcdm\  when combined with constraints derived from observations of the local peculiar velocity field. Indeed, the three MW and M31 analogue pairs appear to exist within filamentary structures at the edge of local voids as traced by galaxy clustering (see Figure 6 of \citealt{Pillepich_2024}), in broad agreement with observations of the local large-scale environment. Throughout this work, we consider the environments formed in TNG50 around these LG-like pairs to be representative of the local Universe within cosmic variance.

\subsubsection{Our Simulated Galaxy Samples}\label{sec:methods_MWsamples}

For each of the three MW and M31 pairs identified by \citet{Pillepich_2024}, we designate the least massive as the MW analogue, motivated by prior observational studies (e.g. \citealt{Klypin_2002, Carlesi2022}). We then define a galaxy sample for each TNG50 MW analogue by transforming the positions and velocities of all other galaxies to be relative to the MW analogue. Hereafter, we use the terms `LG' and `MW' interchangeably when referring to the samples of simulated galaxies around the three MW + M31 analogue pairs. Throughout this work, we separately consider the environments contained within $D < 1 \ \mathrm{Mpc}$, $1 \ \mathrm{Mpc} < D < 10 \ \mathrm{Mpc}$, and $1 \ \mathrm{Mpc} < D < 25 \ \mathrm{Mpc}$ relative to each of the MW analogues. We do this to enable straightforward comparison to observations, because we generally expect differences in completeness in these regimes due to the distance dependence of identifying low-mass galaxies observationally. The maximal radius of $D = 25 \ \mathrm{Mpc}$ is set by the size of the TNG50 simulation box. 

Table \ref{tab:TNG50_MW_properties} details the basic properties of the MW analogues such as stellar mass and the mass of the host halo identified by the FoF algorithm. The two galaxies in pairs `A' and `B' belong to the same host halo as identified by the FoF algorithm, but those in pair `C' do not. We do not demand that the galaxy pairs belong to the same FoF halo, particularly because we employ a different method of assigning galaxy groups (Appendix \ref{sec:appendixGF}).

To ensure a realistic comparison with observations, we define a set of galactocentric coordinates for each MW analogue based on the angular momentum of its stellar particles. A north galactic pole is defined by the direction of net total angular momentum of the MW subhalo stellar disk. This is calculated as the sum of the angular momenta of the stellar particles gravitationally bound to the MW subhalo about its center of mass, restricted to those particles that are located within twice the radius containing half of the total stellar mass ($2\,R_{1/2}$). The box coordinates are then rotated into these new galactocentric coordinates. To mimic the `Zone of Avoidance' (ZoA), which is the region of the sky traced out by the MW where distant galaxies are obscured in visible light, we exclude galaxies located within the area $|b| < 10^\circ$ that is oriented along the plane perpendicular to the direction of net stellar angular momentum.

In order to probe the lower mass end of the simulated galaxy populations, we select galaxies that contain at least 25 stellar particles. We additionally require that all subhalos be of cosmological origin, as indicated in the SUBFIND group catalogs. This is done to exclude clumps of matter formed instead from baryonic processes such as disk instabilities in already formed galaxies, which SUBFIND cannot distinguish a priori from subhalos formed from cosmological structure formation. Throughout this work, the stellar mass of a simulated galaxy is computed as the sum of star particle masses located within $2\,R_{1/2}$.\footnote{Depending on the spatial concentration of the subhalos, this does not necessarily mean that all 25 stellar particles in the lowest-mass subhalos contribute to the mass in $2\,R_{1/2}$.} Across the three MW environments, this stellar particle limit translates to a minimum subhalo mass of $M_\ast(2\,R_{1/2}) \approx 10^6 \, M_\odot$. 

While the low particle limit is not ideal for robust statistical analyses of properties that are sensitive to particle resolution, this selection enables us to characterize the abundances and environmental properties of the low-mass population down to $M_\ast \approx 10^{6} \ M_\odot$. We note that in analyses of color or star formation properties, the region $M_\ast \lesssim 10^{6.7} \ M_\odot$ may not be reliable due to the low number statistics inherent to subhalos with fewer than 100 stellar particles (e.g. \citealt{Donnari2021_quenchedfraction}). We consider the implications of the low particle resolution on our results in \S \ref{discussion:caveats}. This mass range is indicated by a grey-shaded region in all relevant figures. 

\begin{deluxetable}{cccc}\label{tab:Schechter-fits}
\tablecolumns{4}
\tablecaption{Parameters for our best Schechter function fits for each MW analogue in the volumes (\textit{top}) $1 < D/\mathrm{Mpc} < 10$ and (\textit{bottom}) $1 < D/\mathrm{Mpc} < 25$. These were obtained by fitting the binned mass function over the range $10^{6.7} \leq M_\ast/M_\odot \leq 10^{11}$ using a weighted least-squares optimization approach.}
\tablehead{
    \colhead{~} & \colhead{$\log(M^\star/M_\odot)$} & \colhead{$\Phi^\star [10^{-3}\mathrm{Mpc^{-3}}]$} & \colhead{$\alpha$} 
}
\startdata
\cutinhead{$1 < D/\mathrm{Mpc} < 10$}
    MW A & $10.48 \pm 0.18$ & $3.71 \pm 0.90$ & $-1.29 \pm 0.02$\\ 
    MW B & $10.83 \pm 0.26$ & $3.95 \pm 1.22$ & $-1.31 \pm 0.02$\\ 
    MW C & $10.77 \pm 0.18$ & $6.05 \pm 1.33$ & $-1.31 \pm 0.02$ \\
\cutinhead{
$1 < D/\mathrm{Mpc} < 25$}
    MW A & $10.76 \pm 0.13$ & $2.44 \pm 0.38$ & $-1.31 \pm 0.01$\\ 
    MW B & $10.76 \pm 0.13$ & $2.26 \pm 0.35$ & $-1.31 \pm 0.01$\\ 
    MW C & $10.79 \pm 0.12$ & $2.37 \pm 0.34$ & $-1.31 \pm 0.01$ 
\enddata
\end{deluxetable}

\begin{figure*}[ht!]
    \centering
    \includegraphics[width=\linewidth]{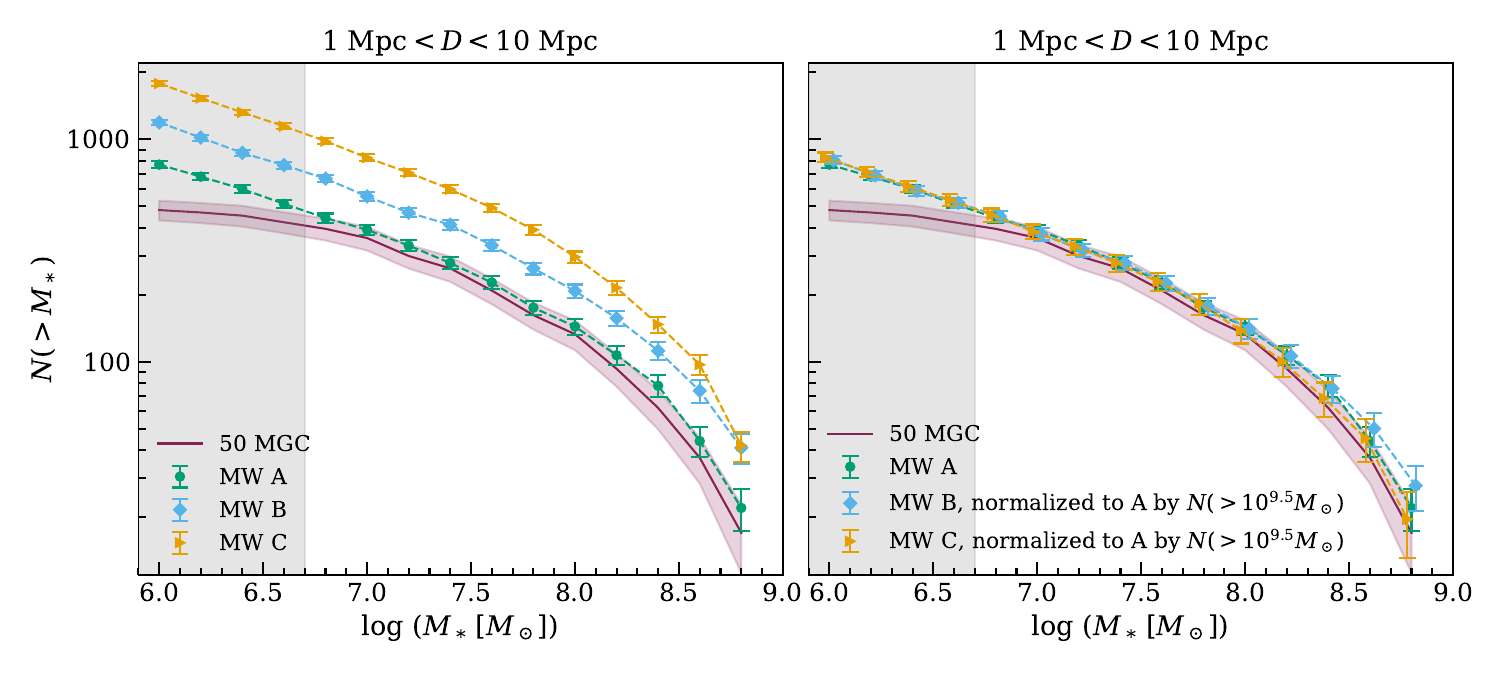}
    \caption{(\textit{Left}) Cumulative number counts (between $10^6 \leq M_\ast/M_\odot \leq 10^9$) in the region $1 <  D/\mathrm{Mpc} < 10$ are shown for the 50 MGC and for each of the three TNG50 MW analogue samples. (\textit{Right}) The cumulative number counts for the MW B and C samples are normalized to those of the MW A sample by the number of bright galaxies, $N(< 10^{9.5} M_\odot)$, in the respective environments within $1 < D/\mathrm{Mpc} < 10$ of each MW analogue. For this figure only, the TNG50 stellar masses are derived from $g-i$ using the TNG photometry. Poisson counting error is shown for the TNG50 counts, and uncertainties in the 50 MGC counts are derived from mass and distance measurement uncertainties in addition to Poisson error. A small offset in $\log M_\ast$ is applied to the MW B and C data points relative to MW A for easier visualization.}\label{fig:DMdensity}
\end{figure*}

\subsection{Observational Data}

We compare our simulated galaxy samples to several observational results, focusing primarily on observed galaxies within $D \leq 25~\mathrm{Mpc}$ for a direct local environment comparison. The optimal choice for this is the 50 Mpc Galaxy Catalog (50 MGC; \citealt{Ohlson_2024}). The 50 MGC assembles homogeneous mass, distance, and morphological information for nearby galaxies from HyperLEDA \citep{HyperLeda2014}, the NASA-Sloan Atlas (NSA; \citealt{Blanton2011}), and the Local Volume Galaxy catalog (LVG; \citealt{Karachentsev2013}). The LVG catalog is nearly complete within $D\sim 8-10~\mathrm{Mpc}$ for galaxies brighter than $M_B \sim -12$ \citep{LVGC2015}. NSA, which is derived from SDSS, covers one-third of the sky and inherits the SDSS main sample photometric depth of $r \approx 22.2$ and spectroscopic depth of $r < 17.77$. The HyperLEDA database \citep{HyperLeda2014} consolidates data from varying sources, offering complete, nearly all-sky coverage of galaxies nominally brighter than $B \sim 18~\mathrm{mag}$ and with typical limiting surface brightnesses of $\mu_B \sim 25~\mathrm{mag/arcsec^2}$ \citep{HyperLeda2003}. Combining these sources yields $15,424$ galaxies in the 50 MGC.

Of all galaxies in the 50 MGC, $11,740$ have cataloged luminosities and measurements for at least one of four colors ($g-i$, $g-r$, $B-V$ and $B-R$), which \citet{Ohlson_2024} used to calculate consistent stellar mass estimates. We use this subset of the catalog for comparison with simulation data. The stellar mass estimates in the 50 MGC are based on the $g-i$ color-stellar mass relation adopted by the Galaxy and Mass Assembly (GAMA) survey \citep{Taylor2011_GAMA}. To account for the varied photometry of the composite sample, \citet{Ohlson_2024} derived color--mass-to-light relations in the remaining three colors, using galaxies with multiple color measurements to ensure self-consistency. Consequently, all $11,740$ galaxies with color measurements were assigned both mass and $g-i$ color estimates. For one object (Mrk 1271), we believe that the mass estimate is artificially inflated by unreliable photometry, so we select a literature value of $M_\ast = 10^{8.28} M_\odot$ \citep{mrk1271_2023}. \citet{Ohlson_2024} assume a Chabrier IMF, so no conversion is needed for comparison with the TNG stellar masses. 

The $g-i$ color estimates derived by \citet{Ohlson_2024} were also used to separate blue and red populations. All galaxies in this sample also have a `best-distance' estimate, which was obtained following a prescription outlined in \citet{Ohlson_2024} based on redshift-based distances corrected for the local velocity field, as well as redshift-independent distances when available. For specific details on how these mass, color and distance estimates were computed, see \citet{Ohlson_2024}. 

\begin{figure*}[ht!]
    \centering
    \includegraphics[width=\linewidth]{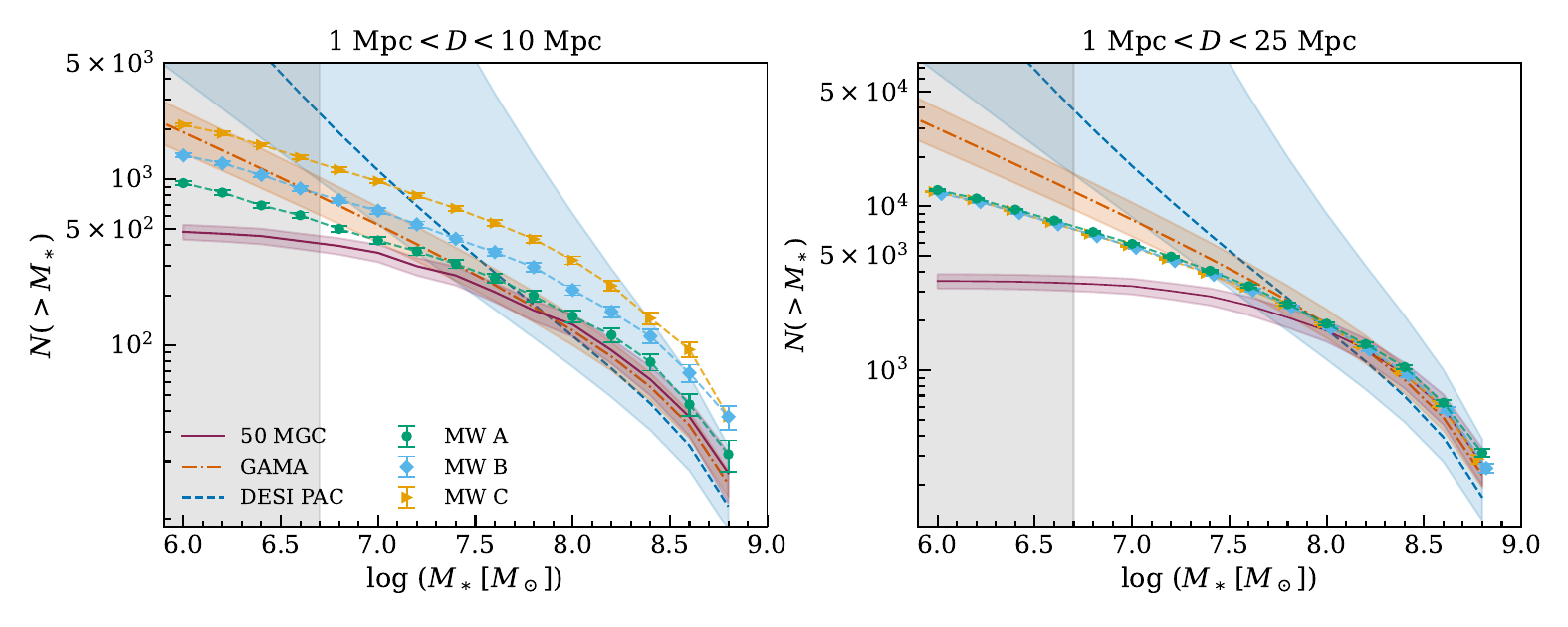}
    \caption{Cumulative number counts (between $10^6 \leq M_\ast / M_\odot \leq 10^9$) as a function of mass for the TNG50 MW samples are compared to observational data, within (\textit{left}) $1 < D/\mathrm{Mpc} < 10$ and (\textit{right}) $1 < D/\mathrm{Mpc} < 25$. For comparison to GAMA Data Release 4, we show the double Schechter function of \citet{Driver_2022} for all survey regions ($z < 0.1$). The $z = 0$ correction and renormalization to SDSS are applied, and the SMF is integrated to produce cumulative number counts within the volume of interest. The triple Schechter fit of \citet{DESI_PAC_2025} to DESI Y1 data is similarly converted to number counts. Error estimates for DESI and GAMA are derived from uncertainties in the fit parameters. The gray region indicates where there are fewer than 100 stellar particles in the TNG50 subhalos. The 50 MGC and TNG50 data are plotted in bins of $\Delta \log (M_\ast/M_\odot) = 0.2$ dex. A small offset in $\log M_\ast$ is applied to the MW B and C data points relative to MW A for easier visualization.} 
    \label{fig:counts-obs}
\end{figure*}

To increase the number of galaxies with mass estimates in the 50 MGC sample, we cross-match the catalog with the Dark Energy Spectroscopic Instrument (DESI) Extragalactic Dwarf Galaxy Catalog,\footnote{\href{https://data.desi.lbl.gov/doc/releases/dr1/vac/extragalactic-dwarfs/}{https://data.desi.lbl.gov/doc/releases/dr1/vac/extragalactic-dwarfs/}} which is derived from the DESI Data Release 1 \citep{DESI_DR1}. We identify an additional 109 galaxies with mass estimates from DESI that do not have mass estimates in the 50 MGC. In the DESI sample, masses are derived from $g-r$ according to the relation of \citet{SAGA2024}, assuming a Kroupa IMF (\citealt{Kroupa2001}), which we correct to a Chabrier IMF. We assume a systematic error of $0.2$ dex for these mass measurements. These galaxies have a median mass of $M_\ast \sim 10^{7.7} \, M_\odot$ and a median distance of $D\sim 33~\mathrm{Mpc}$, and 14 out of the 109 have masses between $10^6 \leq M_\ast/M_\odot \leq 10^7$.

Because the 50 MGC masses were based on GAMA color-mass relations, mass completeness within the 50 MGC sample was approximated based on the GAMA stellar mass functions of \citet{Driver_2022}. \citet{Ohlson_2024} estimate that the sample is complete to a distance of 10 Mpc for masses $M_\ast \gtrsim 10^{7.2} \, M_\odot$, and to a distance of 25 Mpc for $M_\ast \gtrsim \, 10^{7.8} M_\odot$. While group membership information is also available for the 50 MGC sample, based on the \citet{Lambert2020_groupcat} group catalog for the 2MASS redshift survey, we opt to apply our own group finding algorithm to the 50 MGC catalog for consistent comparison with our TNG50 analysis (see Appendix \ref{sec:appendixGF}). When computing group membership, and when presenting all results comparing simulations and observations, we restrict the 50 MGC data to $M_\ast \geq 10^{6} \,M_\odot$ to be consistent with the mass limit of the TNG50 sample.

To compare the TNG50 data to the Local Group observations within $D \leq 1~\mathrm{Mpc}$, we select the Local Volume Database (LVDB; \citealt{Pace2025LVDB}). The LVDB is complete for known dwarf galaxies within $D \sim 3~\mathrm{Mpc}$, making it ideal for studies of the Local Group. We convert the absolute magnitudes provided by \citet{Pace2025LVDB} to stellar masses assuming a mass-to-light ratio of $2$.

We also compare the MW analogue samples with GAMA Data Release 4 (DR 4; \citealt{Driver_2022}) and the recent SMFs derived from the DESI Y1 Bright Galaxies Sample (BGS) by applying the ``Photometric Objects Around Cosmic Webs" method (PAC; \citealt{DESI_PAC_2025}). The PAC method allows for the integration of photometric and spectroscopic surveys by computing the excess projected surface density of photometric objects around spectroscopic objects. The GAMA and DESI PAC samples probe greater volumes -- out to redshifts of $z < 0.1$ and $z < 0.2$, respectively -- and their survey depth is sufficient to constrain the galaxy SMF down to $M_\ast \sim 10^7 \, M_\odot$ and possibly to lower masses, albeit with more uncertainty. We convert the GAMA and DESI Schechter functions to equivalent cumulative number counts as a function of mass for comparison to the TNG50 sample within the volumes of interest.

\section{Results} \label{sec:3}

\subsection{Dwarf Galaxy Number Counts} \label{results:counts}

In Fig. \ref{fig:TNG-counts} we present our theoretical predictions for the cumulative number counts within 10 and 25 Mpc for each of the three MW analogue samples. We find that the mass functions within $1 < D/\mathrm{Mpc} < 10$ and $1 < D/\mathrm{Mpc} < 25$ can be described by a single Schechter function, expressed in $\log M_\ast$ space as:
\begin{equation}
\begin{split}
    \Phi(M_\ast)d \log M_\ast = \ln(10) \Phi^\star
    \left( 10^{\log M_\ast-\log M^\star}\right)^{\alpha+1} \\
    \times \exp\left(-10^{\log M_\ast-\log M^\star}\right) d \log M_\ast
\end{split}
\end{equation}
We fit this functional form to each SMF using weighted non-linear least squares minimization (\verb|scipy.optimize.curve_fit|), accounting for the Poisson counting uncertainty within each uniform bin of size $\Delta \log M_\ast = 0.2$ dex. The Schechter function was fit over the mass range $10^{6.7} \leq M_\ast/M_\odot \leq 10^{11}$ to avoid low-resolution effects at the low-mass end and uncertainties at the high mass end due to the limited volume. In Table \ref{tab:Schechter-fits}, we provide the Schechter fit parameters and their associated uncertainties for each MW sample. We integrate the Schechter fits and display the resulting cumulative number counts in Fig. \ref{fig:TNG-counts} (solid lines). We obtain errors in the cumulative number counts by Monte Carlo sampling of the covariance matrix that was estimated via the least squares optimization. The parameter covariances are assumed to follow a multivariate Gaussian distribution. The resulting shaded envelopes in Fig. \ref{fig:TNG-counts} show the 16th and 84th percentiles after $N = 10^4$ Monte Carlo realizations of the cumulative number counts.

In the bottom panel of Fig. \ref{fig:TNG-counts}, we also show the cumulative fraction of galaxies per MW sample that are located within the Zone of Avoidance ($|b| \leq 10^\circ$), as defined in \S \ref{sec:methods_MWsamples}. These galaxies are removed from the MW analogue galaxy samples in all analyses. Within both 10 and 25 Mpc, the exclusion of the ZoA does not introduce any bias in the mass distribution, since the fraction of galaxies in this region is roughly constant with mass, spanning between $\sim10-20\%$ of the population associated with each MW analogue.

The three samples are nearly identical within counting error at all masses within $1 < D/\mathrm{Mpc} < 25$ (Fig. \ref{fig:TNG-counts}b). This convergence of physical properties over the larger volume may reflect either reduced cosmic variance at this scale or the limited TNG50 box size sampling fewer independent regions. The $25~\mathrm{Mpc}$ spherical volume probed covers nearly half of the TNG50 simulation volume, and more than $10\%$ of galaxies are common among the three samples within 25 Mpc. When we restrict the volume to $1 < D/\mathrm{Mpc} < 10$, the number counts differ among the three MW analogue samples (Fig.~\ref{fig:TNG-counts}a). At these scales, we expect that cosmic variance between MW realizations is a dominant driver of differences in number counts. We find that the MW B and C mass functions appear to scale with the MW A mass function by a constant factor largely independent of mass. Under the assumption that galaxies trace the underlying dark matter density field, it is plausible that the difference in number counts between the three MW analogue samples in 10 Mpc may simply be an artifact of different mean dark matter densities in their respective environments, so that the ratio of the number of galaxies $N$ to the total mass in dark matter $M_\mathrm{cdm}$ is proportional between MW analogue realizations at fixed volume; i.e. $N_\mathrm{A}/M_\mathrm{cdm, A} \propto N_\mathrm{B}/M_\mathrm{cdm, B} \propto N_\mathrm{C}/M_\mathrm{cdm, C}$. 

To test this, we select the brightest galaxies in $1 < D/\mathrm{Mpc} < 10$ as our observational tracers of dark matter density. We define bright galaxies as having masses $M_\ast > 10^{9.5} \, M_\odot$, because in this regime the connection between stellar mass and mass in dark matter is better understood. In the right panel of Fig. \ref{fig:DMdensity}, for MW B and C we scale the cumulative number counts at all masses between $10^6 \leq M_\ast/M_\odot \leq 10^9$ by the constant factor $N_\mathrm{A}(>10^{9.5} \, M_\odot)/N_i(>10^{9.5} \, M_\odot)$, where $i$ labels the relevant MW realization (B or C). We find good agreement between the counts with this scaling applied. We therefore conclude that the difference in the local mean dark matter density between the three MW analogues, as traced by the number of bright galaxies $N(> 10^{9.5} \,M_\odot)$ in 10 Mpc, is responsible for the discrepancy in the counts as a function of mass. Our values for $N(> 10^{9.5} \, M_\odot)$ in 10 Mpc are reported in Table \ref{tab:TNG50_MW_properties}.

In Table \ref{tab:TNG50_MW_properties}, we also provide measurements of the local dark matter density within 10 Mpc of the three MW analogues, defined in this work as $\Omega_{\mathrm{cdm}, \mathrm{eff}} \equiv \rho_\mathrm{cdm}(1 \leq D/\mathrm{Mpc}\leq 10) / \rho_{c,0}$. We estimate $\Omega_{\mathrm{cdm}, \mathrm{eff}}$ from the summed mass of all dark matter particles within $1 < D/\mathrm{Mpc} < 10$ of each MW, and we adopt $\rho_{c,0}$ for the \planck cosmology. The local dark matter densities vary greatly between the three MW realizations, from $\Omega_{\mathrm{cdm}, \mathrm{eff}} = 0.29$ for MW A to $\Omega_{\mathrm{cdm}, \mathrm{eff}} = 0.61$ for MW C. By comparison, observational estimates of the local dark matter density are typically lower. For instance, \citet{LV_DMdensity} estimate a total matter density in $D < 11~\mathrm{Mpc}$ of $\Omega_m = 0.17$ based on measurements of the virial mass in groups and clusters. A discrepancy of this size is not unexpected; given the small volume of the TNG50 box, voids on scales of $R \gtrsim 10~\mathrm{Mpc}$ are unlikely to form. This may suggest that the LV-like environments in TNG50 are denser than the observed LV environment.

In Fig. \ref{fig:DMdensity}, we compare the cumulative number counts between the three TNG50 samples and the 50 MGC within $1 \leq D/\mathrm{Mpc} \leq 10$. For the mass interval $10^{7.2} \leq M_\ast / M_\odot \leq 10^9$, where \citet{Ohlson_2024} estimate the 50 MGC observations to be complete within 10 Mpc, we find that MW A is more representative of the observations on the basis of number counts than either B or C. However, we demonstrate in the right panel of Fig. \ref{fig:DMdensity} that the three MW realizations are identical in 10 Mpc when scaled by the number of bright galaxies $N (> 10^{9.5} M_\odot)$ relative to MW A. For simplicity of presentation, we therefore designate the MW A realization as our reference sample, and include more detailed analyses of MW B and C in Appendix \ref{sec:A2_MWBC}.

\begin{figure}[t]
    \centering
    \includegraphics[width=\linewidth]{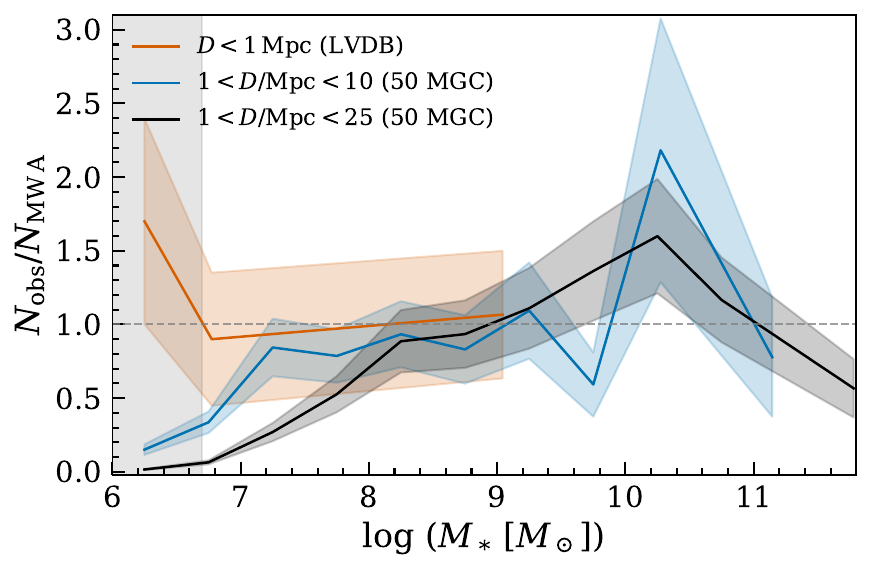}
    \caption{Ratios of galaxy counts ($N_\mathrm{obs}/N_\mathrm{sim}$) within the LVDB and 50 MGC samples relative to the TNG50 MW A sample as a function of mass and distance. Mass bins are determined adaptively to ensure $N_\mathrm{sim}\geq 10$ for reasonable statistics, starting from a width of $\Delta \log(M_\ast/M_\odot) = 0.5~\mathrm{dex}$ and increasing as necessary over the simulation data mass range. This results in wider bins only at smaller distances ($D < 1~\mathrm{Mpc}$) and higher masses ($M_\ast > 10^{11}~M_\odot$). Errors are estimated from the tabulated measurement uncertainties for the observational data and the Poisson error of both the observational and simulation data. The gray dashed line depicts a fraction of unity, equivalent to $100\%$ observational completeness at that mass with respect to the simulated data.}
    \label{fig:completeness_Frac}
\end{figure}

To better illustrate the comparison between the three MW analogues and the observations in Fig. \ref{fig:DMdensity}, we defined the TNG50 stellar masses based on $g-i$ to be consistent with the 50 MGC masses. Using the $g$ and $i$ photometry provided in the TNG50 catalogs and choosing $M_{i, \odot} = 4.53$, we apply the $(g-i)-M_\ast/L_i$ relation of \citet{Taylor2011_GAMA}. Compared to our definition of $M_\ast$ as the stellar mass within $2R_{1/2}$, the color-derived mass estimates skew lower, with an average relative offset of $-0.18 \ \mathrm{dex}$ (Appendix \ref{sec:AppC}, Fig. \ref{fig:mass_rhalf_vs_g-i}).

In Fig. \ref{fig:DMdensity}, and in all analyses of 50 MGC involving galaxy counts, we derive uncertainty estimates for the number counts in each bin by propagating the individual symmetric stellar mass and distance measurement uncertainties reported by \citet{Ohlson_2024} through the mass and distance selection criteria. The upper bound for all mass bins is set at $M_\ast = 10^9~M_\odot$. Each galaxy is treated as a binomial random variable whose probability of contributing to the count in a given mass bin is computed assuming a Gaussian distribution with mean and standard deviation specified by its mass measurement and uncertainty. The sum of the binomial variance of the galaxies gives the total mass-related variance in that bin. An analogous procedure applies for the distance-related component of the total count uncertainty, based on the probability for galaxies in a given mass bin to be located within $1 \leq D/\mathrm{Mpc} \leq 10$ or $1 \leq D/\mathrm{Mpc} \leq 25$. The total variance is the sum of these component mass- and distance-related variances and the Poisson counting variance.

In Fig. \ref{fig:counts-obs}, we compare our theoretical predictions to state-of-the-art observations sourced from the 50 MGC sample and the GAMA DR 4 \citep{Driver_2022} and DESI PAC \citep{DESI_PAC_2025} stellar mass functions. The reported Schechter fits to the DESI PAC and GAMA SMFs are integrated to produce cumulative number counts within $10$ and $25$ Mpc, and uncertainties in the cumulative counts are estimated by Monte Carlo sampling of the Schechter fits within their assumed Gaussian parameter uncertainties. The 16th and 84th percentiles of the resulting distribution of cumulative counts after $N=10^4$ realizations define the shaded envelopes in Fig. \ref{fig:counts-obs}.

We find that the GAMA, DESI PAC, 50 MGC and TNG50 MW A counts agree within error at masses $M_\ast \geq 10^8 M_\odot$ (Fig. \ref{fig:counts-obs}). The trends differ with respect to one another at lower masses within both 10 and 25 Mpc. The cumulative number counts for the 50 MGC sample turn over at lower masses relative to the TNG50 predictions, which may result from mass incompleteness in the observations as depicted in Fig. \ref{fig:completeness_Frac}. Meanwhile, the integrated GAMA and DESI mass functions rise steeply with respect to the TNG50 counts, particularly at masses $10^6 < M_\ast/M_\odot < 10^{7.5}$ (Fig. \ref{fig:counts-obs}). In this regime, the slope of $N(>M_\ast)$ for the MW analogue samples is between $-0.35$ to $-0.36$, compared to a slope of $-0.57$ for GAMA. For DESI, $N(> M_\ast)$ is derived from a triple Schechter fit and does not assume a neatly linear form over this mass range; we estimate a slope of $-1.24$ over the smaller interval $10^6 < M_\ast/M_\odot < 10^7$.

\begin{figure*}
    \centering
    \includegraphics[width=\linewidth]{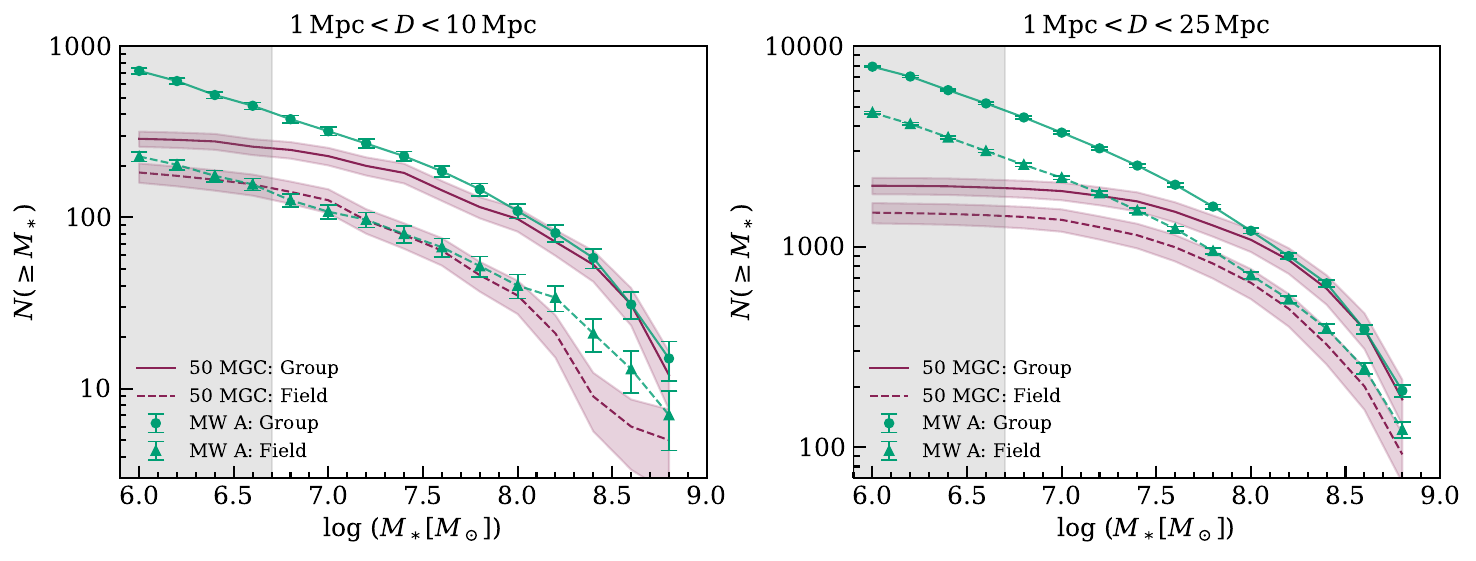}
    \caption{Cumulative galaxy counts between $10^6 \leq M_\ast/M_\odot \leq 10^9$ as a function of field or group classification within (\textit{left}) $1 < D/\mathrm{Mpc} < 10$ and (\textit{right}) $1 < D/\mathrm{Mpc} < 25$, for both the TNG50 MW A simulated galaxy sample and the 50 MGC sample. We define field or isolated galaxies as those which our algorithm classifies as one-member groups.} \label{fig:SMF-env}
\end{figure*}

\begin{figure*}
    \centering
    \includegraphics[width=\linewidth]{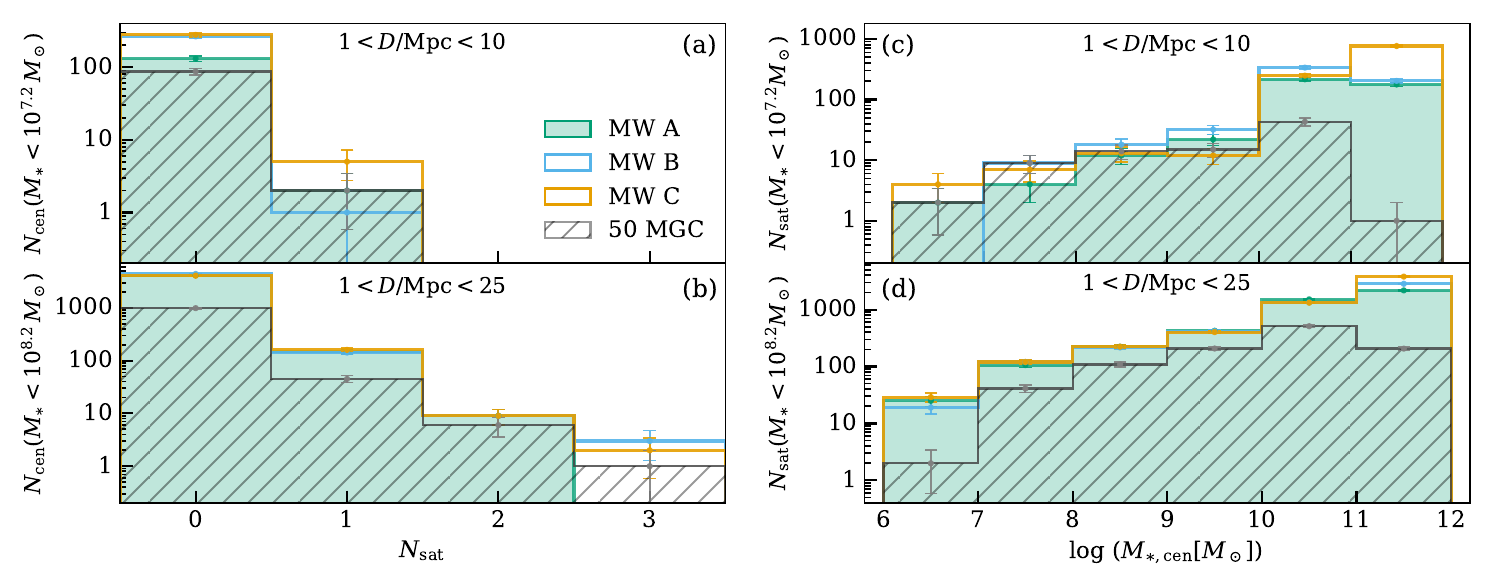}
    \caption{(\textit{Left}) Number of central galaxies with masses below the mass completeness limits of 50 MGC with respect to MW A that have $N_\mathrm{sat}$ satellites within (a) $1 < D /\mathrm{Mpc} < 10$ and (b) $1 < D/\mathrm{Mpc} < 25$. We estimate the mass completeness limits as $M_\ast \leq 10^{7.2} M_\odot$ within $1 \leq D/\mathrm{Mpc} \leq 10$ and $M_\ast \leq 10^{8.2} M_\odot$ within $1 \leq D/\mathrm{Mpc} \leq 25$ (Fig. \ref{fig:completeness_Frac}). Poisson error bars ($\sqrt{N_\mathrm{central}}$) are shown. According to our definition, central galaxies with zero satellites are considered field galaxies. (\textit{Right}) The distribution of satellites below our estimated observational mass completeness limits as a function of the mass of their central within (c) $1 < D /\mathrm{Mpc} < 10$ and (d) $1 < D/\mathrm{Mpc} < 25$. Poisson errors ($\sqrt{N_\mathrm{sat}}$) are shown.}\label{fig:group_analysis}
    \label{fig:Nsat-hist}
\end{figure*}

In Fig. \ref{fig:completeness_Frac}, we compute the ratio of the galaxy count in 50 MGC to the MW A galaxy count within 10 and 25 Mpc as a function of mass. We interpret this as the mass completeness of the observations with respect to the assumed ground-truth of the TNG50 simulation. Uncertainties in the completeness fraction are calculated via analytic error propagation of the Poisson counting error of the TNG50 data and the count uncertainties for the 50 MGC data, which are derived from measurement uncertainties and Poisson error. We estimate that within $1 < D/\mathrm{Mpc} < 10$, the 50 MGC data is incomplete below $M_\ast \lesssim 10^{7.2} M_\odot$, and that within $1 < D/\mathrm{Mpc} < 25$ the mass completeness limit is $M_\ast \lesssim 10^{8.2} M_\odot$. We note a deficit of galaxies with masses $M_\ast > 10^{11} M_\odot$ in 50 MGC relative to MW A. 

For the region $D < 1 \ \mathrm{Mpc}$, we compare the TNG50 data with the Local Volume Database (LVDB; \citealt{Pace2025LVDB}). Magnitudes and distances in the LVDB have asymmetric errors, so the uncertainties on the binned counts are estimated via Monte Carlo sampling of the masses and distances of each galaxy within their assumed skew-normal distributions. Within uncertainty, the LVDB is complete at all masses in 1 Mpc when compared with the TNG50 MW A sample. 

The differences between these two spectroscopic surveys and the TNG50 predictions may be a consequence of the different volumes probed. The GAMA fields extend to $z < 0.1$ \citep{Driver_2022} and the DESI BGS data of \citet{DESI_PAC_2025} extends to $z < 0.2$. It is well known that the MW exists in a Local Void, which begins about 1 Mpc away \citep{TullyFisher1987}, and, according to one estimate, subtends a volume with a width of $R \sim 50-70$ Mpc, depending on the chosen axis \citep{Tully2019_LocalVoid}. The three LG analogues appear to be situated at the edge of voids as traced by galaxy clustering \citep{Pillepich_2024}, although these are relatively small in scale, spanning only a few Mpc. Assuming that MW A is representative of the observable Universe within 25 Mpc, it is plausible that there are fewer galaxies in this volume than predicted by surveys that reach greater volumes due to the presence of a local void. However, we find that $\Omega_\mathrm{cdm} = 0.29$ for MW A within 10 Mpc, which is broadly consistent with cosmological measurements, and at $D < 25 \ \mathrm{Mpc}$, $\Omega_\mathrm{cdm}$ converges on the adopted TNG50 global value of $\approx 0.26$. Additionally, the greater abundance of higher mass ($M_\ast > 10^{11} M_\odot$) galaxies in the TNG50 MW samples relative to 50 MGC, as illustrated in Fig. \ref{fig:completeness_Frac} and Fig. \ref{fig:completeness_BC} (Appendix \ref{sec:A2_MWBC}), does suggest a greater dark matter component in the TNG50 MW analogue samples compared to observations of the local environment. Based on the matter density alone, the TNG50 LG analogues do not appear to exist in a local void at the scales of $10$ or $25$ Mpc. Thus, we cannot strictly attribute the discrepancies in the low-mass end of the stellar mass function between TNG50, DESI and GAMA to environmental differences. 

Throughout this work, we take the TNG50 counts as a raw prediction for the local Universe within $D < 25~\mathrm{Mpc}$. Alternatively, if the GAMA or DESI PAC results are more indicative of the local Universe, the differences at the low-mass end between these two empirical SMFs and the TNG50 count predictions could suggest that many lower mass dark matter halos in the TNG50 MW environments may be devoid of baryonic matter, or that there is a lack of low mass galaxies in the TNG50 simulations. However, the large uncertainties in the DESI PAC results -- as well as the notable discrepancies between GAMA and DESI PAC at the low-mass end -- make it difficult to assess whether either empirical result better represents the low-mass galaxy population within $D < 25~\mathrm{Mpc}$ than TNG50. We elaborate on these points in \S \ref{sec:discussion}. 

\subsection{Counts by Environment}
\label{results:env}
In this section, we explore the distribution of TNG50 galaxies as a function of their environment with the purpose of understanding where the undiscovered population of galaxies may be found. To do so, we separate the TNG50 galaxies according to whether they reside in galaxy groups or in isolation. This requires a definition of ``group" and ``field" that we can consistently apply to observations and simulations. Developing a group finding algorithm which can be used for such a comparison is complicated by several factors; for one, simulations have the advantage of providing full six-dimensional position and velocity information where observations do not. Since we are interested in direct comparisons to observational data, to mimic a typical redshift survey, we reduce our simulation data to two RA- and Dec-like coordinates, a radial distance, and one line-of-sight velocity component. 

Additionally, it is well known that group finding algorithms are affected by survey incompleteness. In Appendix \ref{sec:groupfinder_masscompleteness}, we demonstrate this effect by sampling the simulation data using our estimate of the incompleteness of the 50 MGC defined with respect to the full simulation counts. With this consideration, we apply a method adapted from the algorithm of \citet{Yang_groupfinder_2005} and subsequently \citet{Tinker2022_groupfinder} to classify galaxy groups in both the simulated and observed data. In the \citet{Yang_groupfinder_2005} group finder, group association is dependent on the number density contrast of potential satellites in redshift space. The \citet{Yang_groupfinder_2005} approach is attractive for several reasons: it is physically motivated by the assumption that galaxies trace the underlying dark matter distribution within halos, and furthermore, \citet{Yang_groupfinder_2005} demonstrated that their method may be less subject to contamination from foreground or background galaxies or spurious galaxy associations than FoF group finders. We adapt this approach for our purposes of consistently comparing the simulation and observational data in the presence of incomplete observations and varied photometry. Appendix \ref{sec:appendixGF} details our group finding approach.

In our group finder, centrals and satellites belonging to a group with at least one satellite are considered ``group" galaxies. For both the observed and simulated samples, our algorithm identifies a candidate central, and then probabilistically associates satellites to the central under the assumptions that satellites are most likely to be located within the estimated radius ($R_\mathrm{200}$) of the central's host halo, and that their relative velocities are most likely to lie within three times the halo velocity dispersion ($\sigma_v$). Isolated or ``field" galaxies belong to one-member groups. These are also probabilistically classified such that a galaxy is unlikely to be isolated unless it is at least $>2R_\mathrm{200}$ away from every multi-member group, and it has a relative velocity $>3\sigma_v$. If a preliminary isolated galaxy does not meet this selection criteria, it is  reassigned to the closest multi-member group for which its relative projected distance and line-of-sight velocity are minimal (Appendix \ref{sec:groupfinder_method}). 

Fig. \ref{fig:SMF-env} shows the cumulative number counts for group and field galaxies in the MW A sample compared to the 50 MGC sample after applying our group finder. In the mass-complete regime, estimated to be $M_\ast \gtrsim 10^{7.2}$ within $D < 10$ Mpc and $M_\ast \gtrsim 10^{8.2}$ within $D < 25$ Mpc (Fig. \ref{fig:completeness_Frac}), the relative proportions of field and group galaxies are broadly similar between MW A and 50 MGC. At face value, Fig. \ref{fig:SMF-env} informs us that there is an excess of group and field galaxies in TNG50 compared to the 50 MGC below the observational mass completeness limits, particularly within 25 Mpc. In the remainder of this section, we investigate these discrepancies further. 
 
The TNG50 MW samples predict more isolated galaxies than cataloged in the 50 MGC below the estimated completeness limits. Within 10 Mpc, there are between $1.5-3.2$ times more isolated galaxies in TNG50 depending on the MW realization, and over $4$ times more within 25 Mpc (Fig. \ref{fig:group_analysis}a and \ref{fig:group_analysis}b). Additionally, within $1 < D/\mathrm{Mpc} < 25$, there are approximately four times as many low-mass two-member groups in the TNG50 MW samples for which both the central and satellite masses are less than $M_\ast = 10^{8.2} M_\odot$ (Fig. \ref{fig:group_analysis}b). We note that due to the strict isolation criterion of our group finding algorithm -- which generally translates to a galaxy separation of at least two halo radii away from \textit{every} nearby galaxy group with at least two members (see Appendix \ref{sec:groupfinder_method}) -- many two- to three-member low-mass `groups' may instead be considered \textit{associations} of dwarf galaxies in the field rather than virialized groups.

We also find that satellites with masses below the completeness limit of the 50 MGC observations are overrepresented in TNG50 (Fig. \ref{fig:SMF-env}). We investigate the distribution of these satellites with the mass of their central galaxy in Fig. \ref{fig:group_analysis}c and \ref{fig:group_analysis}d. The greatest disparity between the number of low-mass TNG50 and 50 MGC satellites occurs for satellites associated with the most massive central galaxies ($M_\ast > 10^{11} M_\odot$). It is natural to wonder if this is merely a consequence of the fact that there are more $M_\ast> 10^{11}M_\odot$ galaxies in TNG50 compared to the 50 MGC (Fig. \ref{fig:completeness_Frac} and Fig. \ref{fig:completeness_BC}), and thus potentially more high-mass groups and dense clusters. Indeed, within $1 < D/\mathrm{Mpc} < 10$, there are no multi-member groups with a central galaxy more massive than $M_\ast > 10^{11} M_\odot$ in the 50 MGC, but there are between 2 and 5 such groups in the TNG50 galaxy samples, depending on the MW realization.\footnote{We note that only satellites in Fig. \ref{fig:group_analysis}c and \ref{fig:group_analysis}d are restricted to the specified volume; centrals may be located outside, which is why one low-mass satellite of a more distant central with $M_\ast > 10^{11}~M_\odot$ appears within $10$ Mpc (Fig. \ref{fig:group_analysis}c).}  Within 10 Mpc, there are more than $300$ galaxies with masses $M_\ast \leq 10^9 M_\odot$ that are satellites of either of the two centrals with masses $M_\ast > 10^{11}~M_\odot$ in the MW A sample. This accounts for the majority of the differences in group counts between the MW A and 50 MGC samples in the left panel of Fig. \ref{fig:SMF-env}. For $1 < D/\mathrm{Mpc} < 25$, there are 10 of these high-mass-central groups in the 50 MGC, compared to 25 in the MW A sample and 26 each in MW B and C. In TNG50, these groups have average total satellite counts between $160-260$ depending on the MW realization. These groups contribute a significant portion of satellites below the observational mass completeness limit within 25 Mpc (Fig. \ref{fig:group_analysis}c and \ref{fig:group_analysis}d).

\begin{figure}[t]
    \centering
    \includegraphics[width=\linewidth]{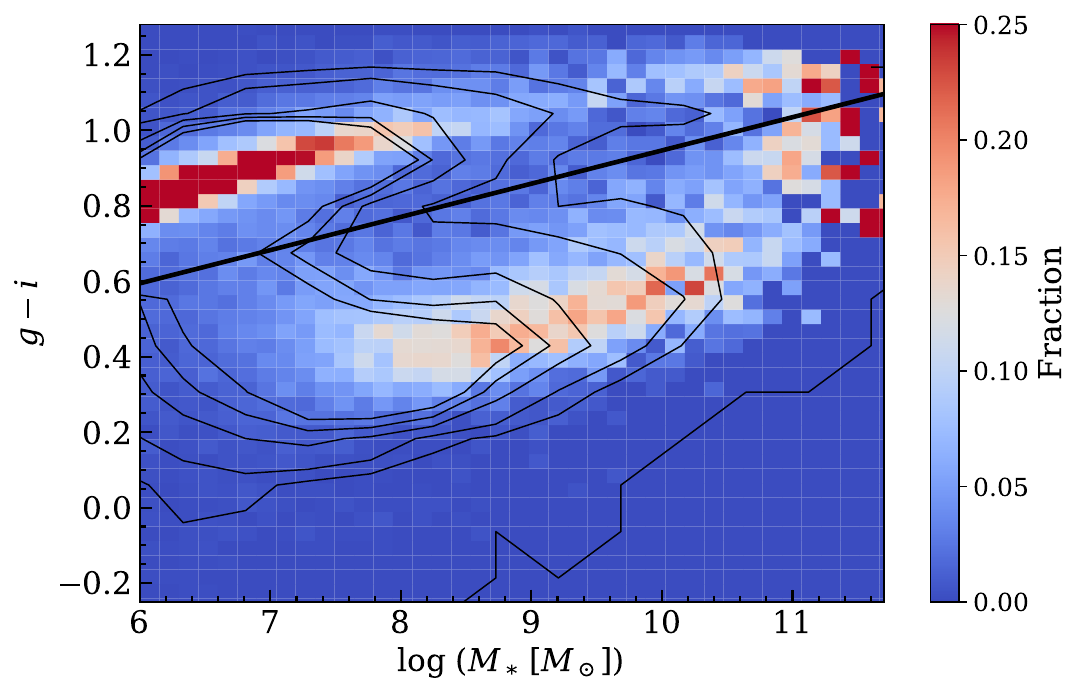}
    \caption{Distribution of all TNG50 subhalos, with at least 25 stellar particles and of cosmological origin, in the plane of $g-i$ vs stellar mass. The line $g-i = 0.09 \log (M_\ast/M_\odot) + 0.07$ was selected to approximate the division between the red and blue population, where blue galaxies lie below this line. To aid in visualization, we normalized each stellar mass bin by the number contained within that bin. Contours represent the unnormalized distribution in levels of 0, 50, and every 100 points thereafter.}
    \label{fig:g_i_mass}
\end{figure}

\begin{figure*}[ht!]
    \centering
    \includegraphics[width=0.98\linewidth]{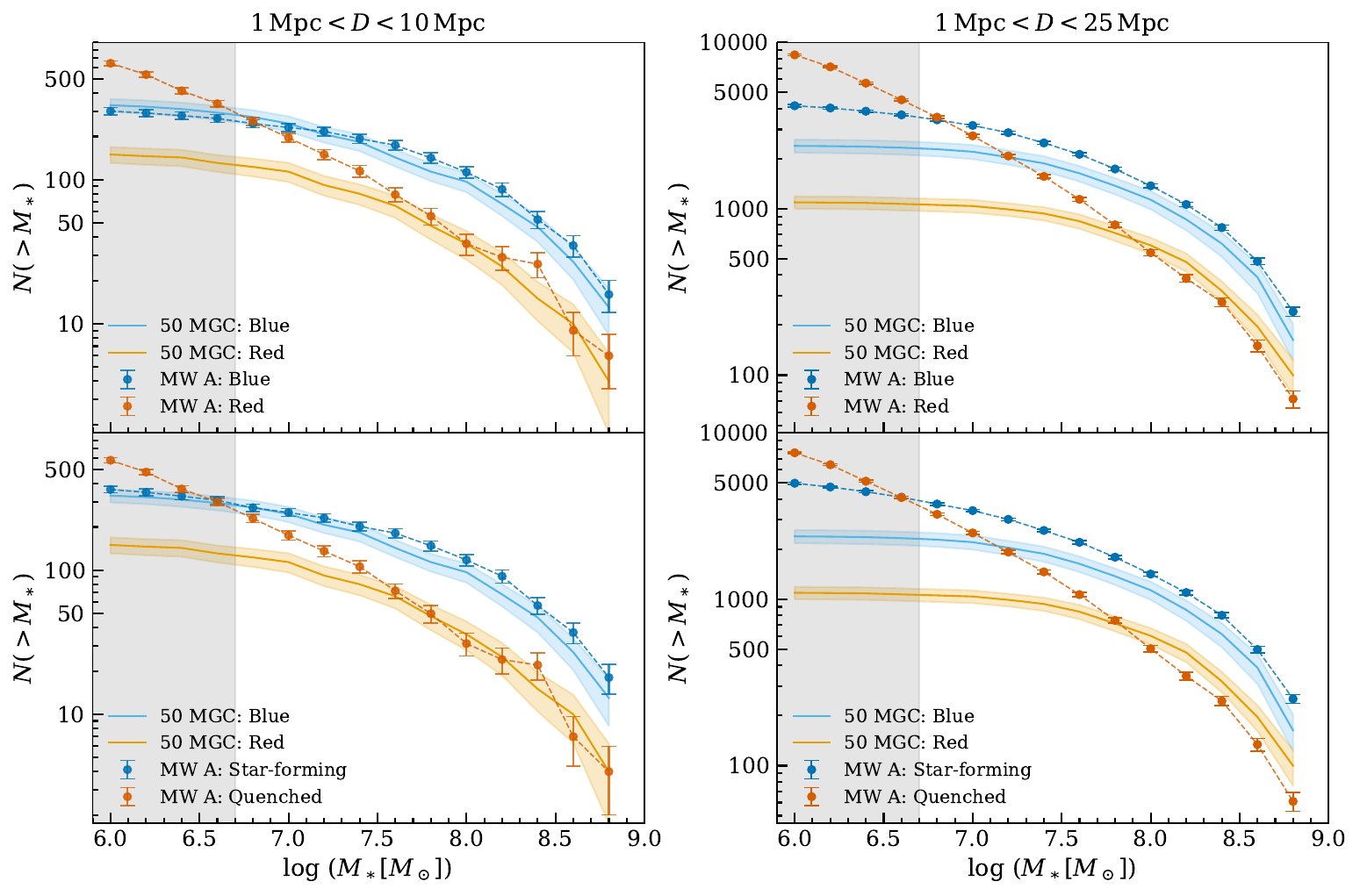}
    \caption{Cumulative galaxy counts between $10^6 \leq M_\ast /M_\odot \leq 10^9$ for the simulated MW A sample selected based on (\textit{top}) $g-i$ color and (\textit{bottom}) position in the SFR$-M_\ast$ plane. The left panels show counts within $1 < D/\mathrm{Mpc} < 10$, and the right panels show counts within $1 < D/\mathrm{Mpc} < 25$. For the TNG50 galaxies, we apply the blue and red separation defined in Fig. \ref{fig:g_i_mass}, and star-forming and quenched criteria as defined in \S \ref{results:SF-color}. The 50 MGC catalog does not provide star formation rates, so the division based on $g-i$ from \citet{Ohlson_2024} is shown in all plots.}
    \label{fig:SMF-color}
\end{figure*}

There may simply be more high-mass structures in TNG50 compared to the nearby Universe in 25 Mpc, as discussed in \S \ref{results:counts} in terms of the total dark matter density, which would make it seem that the observed groups are missing low-mass satellites with respect to the simulated groups. But even at low to intermediate central masses ($10^8 \lesssim M_\ast/M_\odot \lesssim 10^{11}$) within 25 Mpc, which encompasses a region where observations are complete -- and even overabundant -- with respect to the simulated MW samples, we find that there are fewer low-mass satellites (Fig. \ref{fig:group_analysis}d).  
There are $\sim27-29\%$ more multi-member groups within 25 Mpc of the TNG50 MW analogues than we find in the 50 MGC observations at this distance, so it is important that we consider the \textit{average} number of satellites per group that are below the mass completeness limit ($M_\ast < 10^{8.2} M_\odot$), which we abbreviate as $\langle N_{\mathrm{sat},c} \rangle$. For multi-member groups with a central galaxy of mass $10^{8.2} \leq M_\ast/M_\odot \leq 10^{11}$, we find that $\langle N_{\mathrm{sat},c} \rangle \approx 1.5$ for the observations, but $\langle N_{\mathrm{sat},c} \rangle  \approx 3.0$ for the simulations. Thus, relative to the 50 MGC sample, the observed overabundance of the lowest-mass satellites in the TNG50 MW samples -- around low- and intermediate-mass centrals located within $25$ Mpc -- is robust against differences in environmental density.

Meanwhile, within $1 <D/\mathrm{Mpc}<10$, satellite counts below the mass completeness limit are consistent between the observed and simulated samples for central masses $M_\ast \lesssim 10^{10}~M_\odot$ (Fig. \ref{fig:group_analysis}c), suggesting that the greatest disparities between the TNG50 and 50 MGC group counts occur beyond $10~\mathrm{Mpc}$. At any distance, we additionally note that the average number of \textit{all} satellites of all masses -- as a function of the mass of their central galaxy -- is generally consistent between the three TNG50 MW samples and 50 MGC. We therefore conclude that there is a real but modest excess of the lowest-mass satellites in the TNG50 MW samples compared to the 50 MGC observations, and that this excess is most prominent within $D=10-25~\mathrm{Mpc}$. 

\begin{figure*}[ht!]
    \centering
    \includegraphics[width=\linewidth]{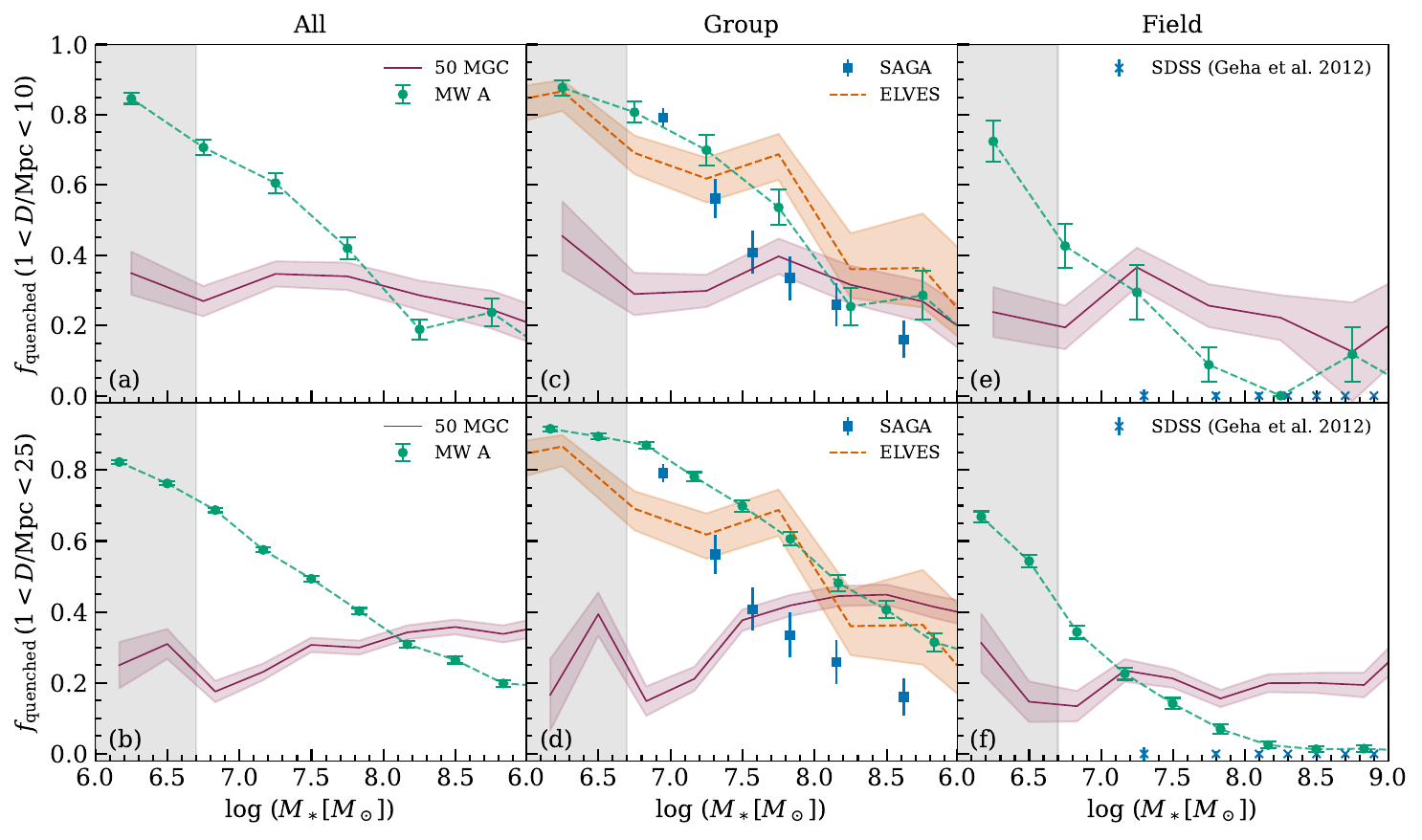}
    \caption{The quenched fraction in the TNG50 MW A sample is compared to the fraction of red galaxies in the 50 MGC sample within (\textit{top row}) $1 < D/\mathrm{Mpc} < 10$ and (\textit{bottom row}) $1 < D/\mathrm{Mpc} < 25$. We also show the results of \citet{SAGA2024} for SAGA satellites, \citet{ELVES_2022} for confirmed ELVES satellites, and \citet{Geha_2012} for SDSS dwarfs in the field. The data in 10 Mpc are plotted in bins of $\Delta \log (M_\ast/M_\odot) = 0.5$ dex, and in 25 Mpc, a mass bin width of 0.33 dex is chosen. (a-b) The quenched and red fractions for all galaxies within the TNG50 MW A and the 50 MGC samples. (c-f) The quenched and red fractions separated by group and field classification. Errors for the TNG50 fractions are computed from binomial variance, and uncertainties for the 50 MGC fractions are derived from measurement and counting uncertainty.}
    \label{fig:quenched_fraction}
\end{figure*}

\subsection{Star-forming vs. Quenched Counts} \label{results:SF-color}

We now investigate the distribution of the simulated low-mass galaxies as a function of color and star formation, as this can inform us as to the number of quenched and therefore faint dwarf galaxies below the completeness limits of current galaxy catalogs. We first consider how to define star formation rates (SFR) for the simulated galaxies, given the resolution limitations of TNG50 and our objective of comparing with observations. We opt to define SFR for the TNG50 data as the sum of the individual SFRs of all gas cells within $2R_{1/2}$. These SFR values are instantaneous, and they lack a direct observational counterpart. To address this, \citet{Donnari_SFR_2019} and \citet{Pillepich_2019} presented SFR measurements computed directly from the stellar particles themselves for subhalos with $\gtrsim 100$ stellar particles. These SFR estimates are averaged over periods ranging from 10 Myr to 1 Gyr. We find that the relative proportion of star-forming and quenched galaxies with $\gtrsim 100$ stellar particles at $z=0$ is broadly similar at all masses regardless of whether we adopt the definition of SFR that is instantaneous, or if we use SFR averaged over 100 Myr. We therefore adopt instantaneous SFRs throughout this work, but with the subtlety that this definition only allows us to identify dwarf galaxies that are quenched at the \textit{current} epoch. This measurement is highly sensitive to rapid variations in global star formation, particularly for galaxies with masses $M_\ast \lesssim 10^{6.7} \ M_\odot$ in which low numerical resolution can artificially increase the stochasticity of star formation, so our analysis at these masses should not be interpreted as capturing the distribution of quenched vs. star-forming galaxies over longer timescales.

We select star-forming and quenched galaxies according to their distance from the $z = 0$ star-forming main sequence (MS) of all TNG50 subhalos with at least 25 stellar particles, in a manner similar to \citet{Donnari_QF_2021}. After performing a linear fit to the MS, we calculate the vertical offset in SFR  ($\Delta\mathrm{SFR_{MS}}=\log({\rm SFR})-\log({\rm SFR_{MS}})$); a galaxy is defined as quenched if $\Delta \mathrm{SFR_{MS}} < -1$ dex and star-forming if $\Delta \mathrm{SFR_{MS}} > -1$ dex. We also note that if quenched galaxies are instead selected according to the threshold $\log (\mathrm{SFR}/M_\ast) \leq -11$, the resulting stellar mass functions, when separated according to their star-forming or quenched designation, are nearly identical.

We additionally consider a color-based separation because the 50 MGC provides colors rather than star formation rates. The TNG SUBFIND catalogs provide synthetic stellar magnitudes for all subhalos in 8 filters. These are intrinsic, rest frame magnitudes based on the sum of the luminosities of all stellar particles gravitationally bound to a subhalo. For background on synthetic photometry in TNG, see \S 3 of \citet{Nelson_2017}.

In Fig. \ref{fig:g_i_mass}, we separate galaxies by color according to their position in the plane of $g-i$ and $M_\ast$; we choose $g-i$ for consistent comparison with the 50 MGC. We perform a linear fit to the red sequence (RS) in the color-mass plane, and compute the vertical offset in $g-i$ ($\Delta_\mathrm{RS}$) for all galaxies. We then translate the RS line by the value of $\Delta_\mathrm{RS}$ where the density of galaxies is minimal between the blue cloud and red sequence. This yields a blue vs. red separation line:
\begin{equation}
    g-i = 0.09\log(M_\ast/M_\odot) + 0.07
\end{equation}
where galaxies lying above are red, and those below are blue. In the TNG50 MW analogue samples, there are no significant differences between our defined red and quenched populations and between our defined blue and star-forming populations (Fig. \ref{fig:SMF-color} and Appendix \ref{sec:A2_MWBC}, Fig. \ref{fig:color-all}).

In Fig. \ref{fig:SMF-color}, we compare the cumulative mass distributions of blue and red galaxies between $10^6 \leq M_\ast/M_\odot \leq 10^9$ in the TNG50 MW A sample and in the 50 MGC. Within 10 Mpc, we observe significantly more red (quenched) galaxies at low masses within the TNG50 sample than in the 50 MGC data. This result is in stark contrast with the blue (star-forming) populations, which show good agreement between TNG50 and the 50 MGC observations across the whole mass range. Within 25 Mpc, the discrepancy between predicted and observed red (quenched) populations grows even further, while the blue (star-forming) SMFs show broad agreement down to $M_\ast \sim 10^{7.5}\ M_\odot$, below which the 50 MGC counts deviate from those predicted by TNG50. This is possibly a byproduct of observational selection biases, whereby blue galaxies are preferentially observed because their active star formation boosts their overall luminosity.

In Fig. \ref{fig:quenched_fraction}, we show the quenched (red) fraction in the TNG50 MW A and 50 MGC samples for all galaxies, and for group and field galaxies separately. The uncertainties for the TNG50 quenched fractions are computed from standard binomial variance, and uncertainties in the 50 MGC fractions include contributions from both measurement and counting uncertainty. For each bin, galaxies are assigned a membership probability weight based on their assumed Gaussian measurement uncertainties, similar to the calculation of cumulative number count uncertainties in \S \ref{results:counts}. The resulting total weighted binomial variance is then added in quadrature with the binomial variance from the deterministic counts to obtain the overall uncertainty. The total quenched fraction in the TNG50 samples rises at low masses, and we note similar results for the B and C samples (Appendix \ref{sec:A2_MWBC}, Fig. \ref{fig:quenched_fraction-all}). However, the fraction of red galaxies as a function of mass in the 50 MGC sample is comparatively flat. This aligns with the expected incompleteness of the red population in 50 MGC at low masses, as can be seen in Fig. \ref{fig:SMF-color}. 

For galaxies in groups, the quenched fraction in the TNG50 MW A sample broadly reproduces observational \citep{ELVES_2022, SAGA2024} and simulation results \citep{mercado2025} for satellites around MW-mass galaxies, notwithstanding differences in volume and environment selection between our work and these studies. \citet{mercado2025} report that for the satellites of over 100 MW analogues in TNG50 (which are not required to be members of a MW + M31 pair), the quenched fraction rises above $0.5$ for masses less than $M_\ast \sim 10^8 M_\odot$, in agreement with our results in Fig. \ref{fig:quenched_fraction}c and \ref{fig:quenched_fraction}d. 

We show in Fig. \ref{fig:quenched_fraction}c and \ref{fig:quenched_fraction}d that the overall trend in the quenched fraction for satellite galaxies in the Satellites Around Galactic Analogs survey (SAGA; \citealt{SAGA2024}) is broadly consistent with the observed trends for TNG50 group galaxies. \citet{SAGA2024} correct for survey incompleteness in their computation of the quenched fraction, which makes the SAGA result more suitable for comparison with simulations than our analysis of the 50 MGC data, in which we do not apply any correction for catalog incompleteness. However, the simulated galaxy samples generally predict a higher quenched fraction for galaxies in groups at all masses compared to the SAGA results, particularly within 25 Mpc. These discrepancies could arise from a number of factors, including differences in the probed environments; our samples include low-mass centrals as well as satellites associated with galaxies of any mass, and SAGA spans a greater volume ($25 \lesssim D/\mathrm{Mpc} \lesssim 40$), so we are not making a direct comparison.

In Fig. \ref{fig:quenched_fraction}c and \ref{fig:quenched_fraction}d we also compare our TNG50 results with the quenched fraction of satellites from the Exploration of Local VolumE Satellites survey (ELVES; \citealt{ELVES_2022}). The ELVES sample consists of $251$ confirmed satellite candidates around MW-mass galaxies within $D < 12~\mathrm{Mpc}$. Based on volume, ELVES may be more aligned than SAGA with our study of the quenched fraction in groups when we restrict the volume to $1 < D/\mathrm{Mpc} < 10$ (Fig. \ref{fig:quenched_fraction}c). \citet{ELVES_2022} report a satellite quenched fraction of $\sim 85\%$ at $M_\ast \approx 10^6 M_\odot$, which is consistent with our result for galaxies in groups within 10 Mpc. At all masses, the overall trend in the quenched fraction of ELVES satellites agrees well with our findings.

For galaxies in the field, we observe a nonzero quiescent fraction in our TNG50 MW analogue samples, which is in contrast with the results of \citet{Geha_2012} for SDSS field galaxies. We find near-consistent agreement only at higher masses ($M_\ast \gtrsim  10^{8.2} M_\odot$). This may be an artifact of our different definition of a field galaxy, particularly because the red fraction for field galaxies in the 50 MGC sample also does not agree with the results of \citet{Geha_2012}. While our classification scheme depends on the estimated virial radii of nearby galaxies, \citet{Geha_2012} classify dwarf field galaxies as those for which the nearest luminous galaxy is a fixed distance of at least 1.5 Mpc away. Alternatively, as considered by \citet{Geha_2012}, the quenched fractions for their sample may be biased if there is a population of quenched, low-surface-brightness galaxies in the field below the detection limits of SDSS.

Although in observations quenched dwarf galaxies are exceedingly rare in the field, some simulations do predict a small but nonzero population of quenched, isolated dwarfs (e.g. \citealt{Fillingham2018, Dickey_2021}). Indeed, we find that the quenched fraction in the field for the TNG50 MW samples is $> 0.2$ for masses $10^{6}-10^7 M_\odot$, reaching $0.7$ at the lowest masses (Fig. \ref{fig:quenched_fraction}e and \ref{fig:quenched_fraction}f). Our result, while focused on the subvolumes of the TNG50 box surrounding three LG analogues, is consistent with other recent studies of the field quenched fraction in TNG50. \citet{Bhattacharyya2025} find that the quenched fraction of TNG50 dwarf field galaxies with masses $10^{7.5} < M_\ast / M_\odot < 10^9$ lies between 0 and 0.2 at $z = 0$. \citet{Benavides_2025} also examine the $z = 0$ field dwarf population in TNG50, but extend their study down to $M_\ast = 10^7 M_\odot$. While they define field dwarfs differently -- as galaxies with masses $10^7 \leq M_\ast/M_\odot \leq 10^9$ that are classified as isolated centrals by the FoF algorithm -- \citet{Benavides_2025} similarly note an upturn in the quenched fraction at low masses.  We confirm this finding and show that the quenched fraction in the field continues to rise at masses $M_\ast < 10^7 M_\odot$.

We finally infer how much of the mass incompleteness in the 50 MGC is biased towards quenched (red) and isolated galaxies, using the simulated galaxy counts as a prior. Below the mass completeness limit of $M_\ast \lesssim 10^{8.2}\,M_\odot$ and within 25 Mpc, the posterior probability that a galaxy missing from the 50 MGC is both quenched (red) and isolated is at least $15\%$. This should be interpreted as a conservative, lower-bound estimate, since we do not account for the hundreds of satellites around centrals with masses $M_\ast > 10^{11}\,M_\odot$ that are found in TNG50 but not in the observed local Universe. This may possibly inflate the total dwarf satellite count as discussed in \S \ref{results:env}, thereby reducing our estimate of the fraction of galaxies missing from current catalogs that are in the field. 

\section{Discussion} \label{sec:discussion}

\subsection{Low-Mass Galaxy Discovery Space}

Future surveys such as LSST, Euclid, and Roman will uncover thousands of new dwarf galaxy candidates within $D < 25~\mathrm{Mpc}$. For this reason, we have undertaken the task of studying the mass distribution and properties of low-mass galaxies in the TNG50 simulation to make predictions for the undiscovered dwarf galaxy population. In doing so, we have compared current best observations of the local dwarf galaxy population with our predictions from TNG50. We believe that the discrepancies in number counts that exist between our predictions and current observations (the 50 MGC) stem from observational selection bias effects resulting in mass incompleteness, and that future surveys such as LSST have the potential to resolve these differences.

Below our estimate of the mass completeness limits of the 50 MGC, we find considerably more isolated galaxies in the TNG50 MW samples than in the observations, both within 10 and 25 Mpc (Fig. \ref{fig:SMF-env} and \ref{fig:group_analysis}). Many dwarf galaxy searches target satellites of more massive groups, since it is generally more difficult to find low-mass and low-surface-brightness galaxies in the field due to surface-brightness selection effects or distance uncertainties. Additionally, we find a non-negligible population of small, low-mass ``groups'' or dwarf galaxy associations in the TNG50 MW samples. All galaxies in these associations -- which typically consist of 2 or 3 members -- have masses below our estimated observational mass completeness limits (Fig. \ref{fig:group_analysis}). The implication of these findings is that \textit{observations of the local Universe are missing many dwarfs that are in isolation or exist in low-density environments}.

At all satellite and group masses, there is no evidence that the average number of satellites per group differs between the simulations and observations, but we do find that the lowest-mass satellites are overrepresented in the TNG50 samples. Within $10-25~\mathrm{Mpc}$, we note that below the estimated mass completeness limit, the average number of satellites around intermediate-mass centrals ($10^{8.2}\leq M_\ast/M_\odot \leq 10^{11}$) is elevated in the TNG50 MW samples relative to observations. This result is robust against differences in the density of the TNG50 MW environments and the observed one (see \S  \ref{results:env}). Our findings suggest that observations are likely missing some of the lowest-mass satellites, particularly beyond the Local Volume where high-sensitivity measurements are difficult to obtain.

A significant majority of TNG50 galaxies below our estimate of the mass completeness limits of the 50 MGC observations are red and quiescent (Fig. \ref{fig:SMF-color}), which suggests that \textit{present observations are predominantly missing quiescent low-mass galaxies}. We interpret the substantial lack of red dwarf galaxies in the observations compared to the simulations as an observational limitation; blue galaxies are easier to observe because their active star formation boosts their total luminosity. When we investigate the distribution of quenched dwarf galaxies by their environment within $10$ and $25$ Mpc of the TNG50 MW analogues (Fig. \ref{fig:quenched_fraction}), we find that our predictions for the quenched fraction of group galaxies generally reproduce the quenched fractions for recent observations of satellites around MW-like hosts \citep{ELVES_2022, SAGA2024}. However, we find a lack of agreement between our predictions and observations for the quenched fraction of field dwarfs (Fig. \ref{fig:quenched_fraction}e and \ref{fig:quenched_fraction}f), which we believe to arise from incompleteness in observations. We therefore conclude that the greatest discovery space for low-mass galaxies exists for quiescent dwarfs in the field. Our result echoes the conclusions of other recent simulation studies that similarly suggest that there is a significant population of quenched dwarf galaxies in isolated environments (\citealt{Bhattacharyya2025, Benavides_2025}). We have expanded upon these works by focusing on the environments around three LG analogues, imposing an element of ``observational realism". Finally, we remark that wide and deep surveys are needed to detect this population, and we explore the implications of our results for LSST in \S \ref{sec:4.2} through \S \ref{sec:4.4}.

The MW analogues in TNG50 appear to exist in denser environments at $10$ and $25$ Mpc than observations suggest, which has implications for the significant quantity of dwarf galaxies in groups in TNG50 compared to the 50 MGC (Fig. \ref{fig:SMF-env}). This is, at least, partially an artifact of the limited box size of TNG50, which does not allow for the formation of large voids. Within 25 Mpc, the value of $\Omega_\mathrm{cdm}$ is consistent with the cosmological average for all MW realizations, and this value grows within 10 Mpc. It is thought that the MW exists at the outskirts of a Local Void extending to at least $\sim60~\mathrm{Mpc}$ (e.g. \citealt{TullyFisher1987, Tully2019_LocalVoid}); empirical estimates of the virial mass in groups and clusters suggest that $\Omega_\mathrm{cdm}$ within the Local Volume should be much lower than the cosmological average (e.g. \citealt{LV_DMdensity}). Within 25 Mpc, there are more than twice as many massive groups with a central of mass $M_\ast \geq 10^{11} M_\odot$ in the TNG50 MW samples compared to observations, which supports the interpretation that the TNG50 MW analogues exist in denser environments. We therefore emphasize that the theoretical estimates presented here are subject to cosmic variance and simulation box size limitations, so that these number counts are possibly an \textit{upper bound} on the true nearby dwarf galaxy count. This is particularly relevant for the red and quiescent count, where the greater number of clusters and groups in the TNG50 volume could result in more environmentally quenched systems.

Ongoing wide and deep surveys such as DESI have led to discoveries of dwarf galaxies in the Local Volume (e.g. \citealt{MartinezDelgado2021, Karachentsev_2022_DESI, Hunter_2025}), and upcoming surveys such as LSST and Roman will further enable resolved and semi-resolved studies of nearby, low-mass galaxies. However, the most complete empirical predictions for the abundance of dwarf galaxies with masses $M_\ast = 10^6-10^7~M_\odot$ come from redshift surveys which probe greater distances. In Fig. \ref{fig:counts-obs}, we show a decisive lack of agreement within 10 and 25 Mpc between GAMA \citep{Driver_2022}, DESI PAC \citep{DESI_PAC_2025}, and the TNG50 predictions in this mass regime. The GAMA and DESI PAC stellar mass functions are not well-constrained at $M_\ast \sim 10^6~M_\odot$, which could contribute to this discrepancy; these redshift surveys also probe volumes out to $z < 0.1$ and $z < 0.2$, respectively, which may not represent the local Universe within 25 Mpc. We assume that the TNG50 simulation results form a ``ground-truth" prediction for the local population of dwarf galaxies, and so within this interpretation, the GAMA and DESI PAC results are inconsistent with the expected abundance of the local population below $M_\ast \lesssim 10^7 M_\odot$. Future observations are needed to empirically constrain the faint end of the local galaxy SMF. Below, we discuss the implications of our results for upcoming wide and deep surveys and discuss the potential caveats to our results.

\subsection{Implications for Flux-Limited Surveys}\label{sec:4.2}

From the standpoint of TNG50, nearly all Local Volume galaxies in the mass range $10^6 \leq M_\ast /M_\odot \leq 10^9$ should be theoretically detectable by current and future surveys; here, we focus on LSST \citep{LSST2019}, which is expected to publish its first data release in the coming year.\footnote{\hyperlink{https://www.lsst.org/about/project-status}{https://www.lsst.org/about/project-status}} We leave to other studies (e.g. \citealt{Mutlu-Pakdil_2021, Tsiane2025}) more detailed analyses involving the expected detectability of dwarf galaxies that will be possible with LSST, and instead make first-order estimates. 

Our analysis of the three LG environments in TNG50 suggests that all galaxies with masses $M_\ast \geq 10^6~M_\odot$ within 10 Mpc -- including quenched systems -- are theoretically detectable with LSST on the basis of the co-added imaging depth of $r < 27.5$. Within 10 Mpc, the faintest $r$-band apparent magnitude is $r = 20.23$ across our three MW samples. Even if we were to consider dust attenuation, this would still be well within the LSST ten-year survey depth. 

\subsection{Spatial Resolution}\label{sec:4.3}

We postulate that the majority of $M_\ast \geq 10^6~M_\odot$ galaxies within 10 Mpc will be spatially resolved, which has implications for resolved and semi-resolved studies. We base our estimate of spatial resolvability on the apparent sizes of the galaxies in our TNG50 MW samples. We approximate the $r$-band half-light radii ($R_{e,r}$) for the galaxies in our samples based on the \citet{TNGStellarSizes2018} catalog, considering both the face-on and smallest edge-on 2D projections. This catalog provides projected half-light radii only for subhalos with at least 100 stellar particles, so we calibrate a relation between $R_{e,r}$ and the half-mass radius ($R_{1/2}$). The scatter around this relation is $
\sigma\sim~360-460~\mathrm{pc}$ for galaxies with $R_{e,r} \leq 1~\mathrm{kpc}$, depending on whether the projection is face-on or edge-on, so our estimates of the angular size of the smallest galaxies in our sample from the reference point of the MW analogue are uncertain by $\sim 7.5''-9.5''$. With this caveat, we find that fewer than 10 dwarf galaxies within $10~\mathrm{Mpc}$ of our three MW analogues have angular widths less than $\sim 15''$. This suggests that nearly all of the expected number of classical/bright dwarf galaxies within $10~\mathrm{Mpc}$, as predicted by the TNG50 MW A sample, may be spatially resolved or semi-resolved by wide and deep surveys such as LSST, assuming the angular resolution is atmosphere-limited or better ($\sim 0.7''$). 

The TNG simulations have successfully reproduced the galaxy size-mass relation within observational uncertainties at various masses \citep{TNGStellarSizes2018, RodriguezGomez2018, HuertasCompany2019}, but this is not guaranteed for the low-particle-resolution subhalos considered in our samples, with masses $M_\ast \lesssim 10^{6.7}~M_\odot$. TNG50 has a median spatial resolution of $\sim 100~\mathrm{pc}$ \citep{Pillepich_2019, Nelson_2019_TNG50results}, which may introduce numerical resolution effects in estimates of the physical size of the smallest galaxies. Thus, we caution that the above analysis of the spatial resolvability of the TNG50 galaxies in our samples could be affected by the resolution limits of the simulation, and so it should be interpreted as a rough estimate.

\subsection{Sky Coverage}\label{sec:4.4}

The variation in local environment as a function of line-of-sight can have real implications for surveys of the nearby dwarf galaxy population, where local density fluctuations dominate over cosmological homogeneity. This can limit galaxy statistics for wide-field, ground-based surveys such as LSST, which has a footprint of $18,000~\mathrm{deg}^2$ covering $\sim44\%$ of the sky. To quantify this, we define a north and south celestial pole relative to our defined north and south galactic poles as described in \S \ref{sec:methods_MWsamples}, and restrict the TNG50 MW samples to two analogous volumes symmetric about these poles. We find that the cumulative number counts presented here can vary by as much as a factor of $\sim 2$ between hemispheres within $10$ and $25~\mathrm{Mpc}$ depending on the MW realization. This suggests that without full-sky coverage, cosmic variance is a significant limiting factor in building a census of nearby low-mass galaxies. Finally, we count the number of galaxies in the MW A realization lying within the LSST footprint -- in either the northern or southern hemispheres, since there is no preferred hemisphere within the simulations -- and subtract the number of 50 MGC galaxies located within the LSST footprint in the Southern sky. From this, we estimate that LSST could discover as many as $\sim 200-300$ classical dwarf galaxies in the Southern sky in $10$ Mpc, and as many as $\sim 5,000-6,000$ in $25$ Mpc.

\subsection{Caveats} \label{discussion:caveats}

Throughout this work, we have assumed that the environments around the three LG analogues in TNG50 form a ``ground-truth" prediction for the local galaxy population. While this is a physically motivated assumption, based on previous findings that the formation of the Local Group is closely linked to its large-scale environment (e.g. \citealt{Forero-Romero_2015, Sorce2016, Carlesi2016, Zhai2020}) and on the success of the IllustrisTNG model in reproducing the fundamental properties of observed galaxies \citep{Nelson_2017, Pillepich_2017_TNGintrostellarcontent, Springel_2017, Marinacci_2018, Naiman_2018}, we must consider the limitations of the TNG model and the caveats that may complicate this picture. 

We have defined simulated galaxies as having $\geq 25$ stellar particles, which translates to a stellar mass of $M_\ast \lesssim 10^{6.7} M_\odot$ within $2R_{1/2}$. This is lower than the usual limit of $\sim 100$ stellar particles imposed in similar studies. One possibility, as we alluded to in \S \ref{results:counts}, is that the stellar mass function for the TNG50 galaxies may be suppressed at these low masses relative to empirical results such as DESI PAC \citep{DESI_PAC_2025} or GAMA DR 4 \citep{Driver_2022} due to numerical effects or model choices rather than physical reasons or observational uncertainties (Fig. \ref{fig:counts-obs}). \citet{Pillepich_2017_TNGintrogalaxyformation} reported suppression of the SMF at $z=0$ in lower resolution runs of their fiducial TNG model. However, even in the highest resolution run, the galaxy SMF at $z=0$ was somewhat suppressed at the low-mass end ($M_\ast \sim 10^8 M_\odot$) compared to predictions from SDSS or previous GAMA data releases. Although \citet{Pillepich_2017_TNGintrogalaxyformation} remark that this difference could simply arise from cosmic variance rather than from model choices, additional suppression at lower masses due to model prescriptions -- potentially compounded by numerical resolution limits -- is a possibility. The observational constraints used to calibrate the TNG model extend down only to $M_\ast \sim 10^8 M_\odot$, so the SMF at lower masses is a prediction of the model and could in principle be affected by uncertainties due to model choices. For instance, the TNG simulations implement stronger and faster stellar winds than the original Illustris model \citep{Vogelsberger2013, Torrey2014}. Although this prescription was introduced to improve the modeling of the galaxy SMF -- which was previously over-predicted by Illustris \citep{Pillepich_2017_TNGintrogalaxyformation} -- the enhanced winds could prevent stellar mass growth and thereby yield too few low-mass galaxies at $z = 0$. But just as the galaxy frontier below $M_\ast \lesssim 10^{6.7} M_\odot$ is uncertain in TNG50, the GAMA DR4 SMF is also not well constrained for $M_\ast < 10^{6.75} M_\odot$, and there are sizable uncertainties in this mass range for the DESI PAC SMF (Fig. \ref{fig:counts-obs}). Some caution is warranted in interpreting both observational and simulated galaxy abundances in this mass regime.

Another potential and related limitation concerns the star-formation properties and derived colors of low-mass galaxies in TNG that have fewer than $100$ stellar particles. Although it is exciting to extrapolate the predicted quenched fraction to lower masses than examined by previous studies (Fig.~\ref{fig:quenched_fraction}), this comes with the important caveat that star formation and color properties may be unreliable below $M_\ast \lesssim 10^{6.7} \, M_\odot$, where particle resolution is limited \citep{Donnari2021_quenchedfraction}. The low resolution inherent at these low masses artificially makes star formation appear more stochastic, which could spuriously increase the quenched population by making these systems much more sensitive to stellar feedback. In fact, \citet{Joshi2021} demonstrate that for galaxies with masses $10^7 \leq M_\ast/M_\odot \leq 10^{10}$ that have no satellites, the quenched fraction across different resolution runs of TNG50 only converges around a mass of $M_\ast \sim 10^{9.5} M_\odot$; at lower masses, lower resolution artificially inflates the quenched fraction. On the other hand, the quenched fraction for satellite galaxies and dwarf group members converges at masses as low as $M_\ast \approx 10^7 M_\odot$ \citep{Joshi2021}, which is encouraging for our results for the quenched fraction in groups and for the overall quenched fraction, which is dominated by group galaxies (Fig. \ref{fig:SMF-env}).  We additionally note that our use of instantaneous SFR may overestimate the quenched fraction at low-masses due to the inherent burstiness of dwarf galaxy star formation, which leads to temporary quiescent periods on the order of tens to hundreds of Myr, as seen observationally and reproduced in simulations (e.g. \citealt{ Weisz_2012, Hopkins2014}). A more adequate assessment of quenching in an absolute sense would involve SFR averaged over longer times. 

We finally note that just as predictions based on cosmological simulations can inform observing strategies and assist in interpreting observations of low-mass galaxies, observational results also motivate improvements to the subgrid models and calibration schemes that underpin modern cosmological simulations of galaxy formation \citep{Vogelsberger2020}. The fleet of upcoming and ongoing wide and deep surveys such as LSST has the potential to significantly improve constraints on the $z=0$ galaxy SMF at masses below $M_\ast\sim 10^8~M_\odot$ and also improve our understanding of stellar feedback and evolution in low-mass, low-metallicity galaxies. As cosmological hydrodynamical simulations advance into the frontier of both high resolution and large-volume statistics (e.g. \citealt{Feldmann2023, VanNest2023}), leveraging improved observational constraints on the mass distribution of low-mass galaxies and their star formation and feedback processes will be critical for updating our theoretical understanding of galaxy evolution.

\section{Conclusions}\label{sec:conclusion}

We have employed the highest-resolution run of the IllustrisTNG suite of hydrodynamical cosmological simulations (TNG50) to establish theoretical predictions for the number of classical and bright dwarf galaxies ($10^6 \leq M_\ast/M_\odot \leq 10^9$) within $D < 25~\mathrm{Mpc}$ of three Local Group analogues, which were identified by \citet{Pillepich_2024}. For consistent comparison to observations, we defined a set of galactocentric coordinates relative to the net stellar angular momentum of each Milky Way analogue. Accordingly, a region akin to $|b|\leq 10^\circ$ was cut out from each sample volume to mimic the Zone of Avoidance. Compared to the 50 Mpc Galaxy Catalog \citep{Ohlson_2024} -- a thorough and homogenized collection of galaxies from HyperLEDA \citep{HyperLeda2014}, the NASA-Sloan Atlas \citep{Blanton2011}, and the Local Volume Galaxy catalog \citep{Karachentsev2013} -- we approximate the completeness of current observations in this mass regime as $\sim 50\%$ within $1 \leq D/\mathrm{Mpc} \leq 10$ and $\sim 25\%$ within $1 \leq D/\mathrm{Mpc} \leq 25$. We note that our number count estimates may represent an upper bound given that the TNG50 MW environments do not appear to exist within local underdensities to the extent that current observations imply. 

Below, we summarize our main findings:

\begin{enumerate}[label=\roman*.]
    \item  In our reference sample, there are $\sim 1,000$ and $\sim 12,000$ galaxies of masses $10^6 \leq M_\ast/M_\odot \leq 10^9$ within 10 and 25 Mpc of the simulated MW analogue, respectively (Fig. \ref{fig:TNG-counts}), compared to $\sim500$ and $\sim 3,500$ observed galaxies within these volumes that have cataloged mass and distance estimates (Fig. \ref{fig:counts-obs}). 
    
    \item We estimate that within $1 < D/\mathrm{Mpc} \leq 10$, current observations are incomplete below $M_\ast \lesssim 10^{7.2} \, M_\odot$ ($M_r \gtrsim -14$ and $M_r \gtrsim -13$ for star-forming and quenched galaxies, respectively). Within $1 < D/\mathrm{Mpc} \leq 25$, observations are incomplete below $M_\ast \lesssim 10^{8.2} \,M_\odot$, or $M_r \gtrsim -16.5$ and $M_r \gtrsim -15$ for star-forming and quenched galaxies (Fig. \ref{fig:counts-obs} and \ref{fig:completeness_Frac}).
    
    \item In our TNG50 MW analogue environments, the overwhelming majority of the simulated galaxies below the mass completeness limits of current observations are red and currently quenched (Fig. \ref{fig:quenched_fraction}). This trend persists within 10 and 25 Mpc.

    \item There is a significant population of low-mass galaxies in isolated and low-density environments in the simulated MW samples that lie below the mass completeness limits of the observations (Fig. \ref{fig:SMF-env} and \ref{fig:group_analysis}). Accounting for differences in density between the observed local environment and the TNG50 MW environments, we find a modest excess of satellites below the observational mass completeness limits in TNG50 compared to the 50 MGC around intermediate-mass centrals ($10^{8.2} \leq M_\ast/M_\odot \leq 10^{11}$), and this is more pronounced within $10-25~\mathrm{Mpc}$.
    
\end{enumerate}

\textit{We propose that the most promising discovery space for dwarf galaxies is quiescent galaxies in the field}. Future wide and deep surveys, such as the Vera C. Rubin Observatory's LSST \citep{LSST2019}, will enable discoveries of many such objects that have historically been elusive in observations. We believe that Rubin will be capable of detecting and spatially resolving the majority of Local Volume galaxies with masses $10^6 \leq M_\ast \leq 10^9$ within the LSST footprint, and that it may uncover up to $\sim 5,000-6,000$ dwarf galaxies within 25 Mpc. We are at the cusp of a new era for dwarf galaxy studies in the local Universe, and so leveraging the simulation results presented here can assist in searching for nearby low-mass galaxies and interpreting the results. In turn, these observational endeavors can provide empirical insights into the physical processes that guide our theoretical understanding.

\begin{acknowledgments}
We thank David Ohlson and Anil Seth for helpful discussions related to the 50 Mpc Galaxy Catalog, and Hien Nguyen for assistance with catalog cross-matching.

MBK acknowledges support from NSF grant AST-2408247; NASA grant 80NSSC22K0827; HST-GO-16686, HST-AR-17028, JWST-GO-03788, and JWST-AR-06278 from the Space Telescope Science Institute, which is operated by AURA, Inc., under NASA contract NAS5-26555; and from the Samuel T. and Fern Yanagisawa Regents Professorship in Astronomy at UT Austin.

This research used data obtained with the Dark Energy Spectroscopic Instrument (DESI). DESI construction and operations is managed by the Lawrence Berkeley National Laboratory. This material is based upon work supported by the U.S. Department of Energy, Office of Science, Office of High-Energy Physics, under Contract No. DE–AC02–05CH11231, and by the National Energy Research Scientific Computing Center, a DOE Office of Science User Facility under the same contract. Additional support for DESI was provided by the U.S. National Science Foundation (NSF), Division of Astronomical Sciences under Contract No. AST-0950945 to the NSF’s National Optical-Infrared Astronomy Research Laboratory; the Science and Technology Facilities Council of the United Kingdom; the Gordon and Betty Moore Foundation; the Heising-Simons Foundation; the French Alternative Energies and Atomic Energy Commission (CEA); the National Council of Humanities, Science and Technology of Mexico (CONAHCYT); the Ministry of Science and Innovation of Spain (MICINN), and by the DESI Member Institutions: www.desi.lbl.gov/collaborating-institutions. The DESI collaboration is honored to be permitted to conduct scientific research on I’oligam Du’ag (Kitt Peak), a mountain with particular significance to the Tohono O’odham Nation. Any opinions, findings, and conclusions or recommendations expressed in this material are those of the author(s) and do not necessarily reflect the views of the U.S. National Science Foundation, the U.S. Department of Energy, or any of the listed funding agencies.
\end{acknowledgments}

\vspace{5mm}

%\facilities{}

\software{ Aboria \citep{aboria}, Astropy \citep{astropy:2013, astropy:2018, astropy:2022}, Colossus \citep{colossus},
Matplotlib \citep{matplotlib}, Numpy \citep{numpy}, Scipy \citep{SciPy}}

\bibliography{citations}{}
\bibliographystyle{aasjournal}

\appendix 

\section{The Group Finding Algorithm}\label{sec:appendixGF}
We outline our procedure for assigning galaxy groups, both for the 50 MGC data and the TNG50 simulation data.
In Appendix \ref{sec:A1}, we show that group assignments in a simple halo-based group finder depend on mass completeness, leading to inconsistent results. In Appendix \ref{sec:groupfinder_method}, we describe the group finding algorithm we used in our results presented in the main text. The algorithm presented here for grouping galaxies is published and available at \href{https://github.com/evangelashread/obs-sim-halo}{https://github.com/evangelashread/obs-sim-halo}.

\subsection{A Simple Halo-Based Group Finder}\label{sec:A1}

In this group finding approach -- as in the algorithm we ultimately use in our analysis (Appendix \ref{sec:groupfinder_method}) -- we assign halo masses to galaxies in a manner reminiscent of rank-order abundance matching in simulations (e.g. \citealt{GalaxyHaloConnection2018}). The algorithm is based on the simplifying assumptions that the central galaxy in a group is the most massive and that it is located at the physical center of its host halo. We presume that halos are likely to be in virial equilibrium such that most satellites are located within one halo radius ($R_{200}$) of the group to which they are associated, but this is not a strict requirement -- we allow for the possibility of satellites within $2 R_\mathrm{200}$, where backsplash galaxies may reside (e.g. \citealt{Gill2005}). The velocity condition for satellites is set at 3 times the assumed line-of-sight velocity dispersion for the halo ($\sigma_v$), and a galaxy outside the halo radius, but within 2$R_{200}$, may be classified as a satellite if it does not meet our criteria for classification as an isolated galaxy. Galaxies are considered one at a time from most massive to least, and an empirical SHMR is used to connect the TNG50 galaxy stellar masses to their expected host halo mass. 

In analyzing the simulations, we consider two approaches: a 6-dimensional implementation which leverages the full information available from 3 position and 3 velocity components, and an ``observationally consistent" implementation which only makes use of projected (great-circle) distances and a line-of-sight velocity component. The latter implementation is used for direct comparisons to observation, but the full 6D approach is useful for assessing the performance of our observationally consistent algorithm.

In this algorithm, galaxies are selected iteratively from the set of all ungrouped subhalos ordered by descending $M_\ast$; thus, halos are identified from the most massive to the least. In simulations, we only consider subhalos which are of cosmological origin and have at least 25 stellar particles. We utilize the entire volume of the simulation box, excluding the Zone of Avoidance defined for each MW sample ($|b| \leq 10^\circ$). In all virial calculations we assume $z = 0$ because corrections due to redshift within the 50 Mpc box volume are almost negligible.

\subsubsection{Observationally Consistent Approach} \label{sec:A1_obs}
The outline of the ``observationally consistent" implementation is as follows:

\textbf{Part 1}. The most massive, ungrouped galaxy is selected and then assigned an estimate of the mass of its host halo ($M_h$) from the SHMR relation of \citet{Behroozi_2019}. An estimate of the halo radius ($R_{200}$) is obtained from $M_h$, where $R_{200}$ is defined as the radius at which the halo has an average over-density of $200$ times the critical density of the Universe at $z = 0$ ($\rho_{c,0}$):
\begin{equation}
    R_{200} = \left[\frac{M_h}{(4\pi/3) 200 \rho_{c,0}}\right]^{1/3}
\end{equation}
The virial velocity is computed as $V_\mathrm{vir} = (GM_h / R_{200})^{1/2}$, and we assume a line-of-sight velocity dispersion $\sigma_v = V_\mathrm{vir}/\sqrt{2}$.
    
For a given central galaxy, candidate satellites are then identified. A galaxy is preliminarily assigned as a satellite of the central if all three criteria are met:
\begin{enumerate}[label=\roman*.] 
    \item The candidate satellite has not already been assigned to a group.
    \item The great-circle distance ($d$) between the satellite and central is less than $R_{200}$, with $d$ defined by:
    \begin{equation}
        d = D_\mathrm{cen} \tan \theta ,
    \end{equation}
    where:
    \begin{eqnarray}
        \tan \theta = \frac{2\sqrt{\mathrm{hav}(\theta) (1-\mathrm{hav}(\theta))}}{\cos \theta} \\
        \cos \theta = \frac{\mathbf{D}_\mathrm{cen} \cdot \mathbf{D}_\mathrm{sat}}{D_\mathrm{cen} D_\mathrm{sat}} \\
        \mathrm{hav}(\theta) = \frac{1-\cos \theta}{2}
    \end{eqnarray}
    The quantities $\mathbf{D}_\mathrm{cen}$ and $\mathbf{D}_\mathrm{sat}$ define the 3D galactocentric coordinates of the central and satellite, and $D_\mathrm{cen}$ and $D_\mathrm{sat}$ are the vector magnitudes equivalent to the line-of-sight distances.
    
    \item The satellite's total line-of-sight velocity relative to the central's line-of-sight velocity is less than $3 \sigma_v$, i.e.
    \begin{equation}
        |v_\mathrm{LOS}^\mathrm{sat} - v_\mathrm{LOS}^\mathrm{cen}| \leq 3 \sigma_v
    \end{equation}
    where for any given galaxy, $v_\mathrm{LOS}$ is computed from:
    \begin{equation}
        v_\mathrm{LOS} = \mathbf{v}_\mathrm{pec}\cdot\frac{\mathbf{D}}{D} + H_0 D
    \end{equation}
    with $H_0 = 67.74 \ \mathrm{km/s/Mpc}$.
\end{enumerate}
    
By default, all ungrouped galaxies are less massive than the current central, so all satellites are guaranteed to be less massive than their central. If no suitable satellites are found, the central is preliminarily classified as an isolated galaxy and assigned to a one-member group. This process is continued until all galaxies have been classified.

\textbf{Part 2}. Because groups are classified from most massive to least, it may be the case that a satellite meets the distance and velocity criteria for more than one halo but is automatically assigned to the most massive. To address these edge cases, we iterate through all satellites to identify the central for which the relative distance and velocity are smallest.

The candidate centrals are filtered to the subset for which $d \leq R_{200}$ and $\Delta v = |v_\mathrm{LOS}^\mathrm{sat} - v_\mathrm{LOS}^\mathrm{cen}| \leq 3 \sigma_v$, where $R_{200}$ and $V_\mathrm{vir}$ are the virial parameters for the host halo of the candidate central. We additionally enforce the requirement that any candidate central be more massive than the satellite. From this subset, the central for which the quantity $\sqrt{(d/2R_{200})^2 + (\Delta v / 3\sigma_v)^2}$ is minimal is selected. A satellite is said to be reclassified if this central is different from the one to which it was previously assigned.

\textbf{Part 3}. In reality, many halos are not in total virial equilibrium, and there may be splashback or infalling satellites outside of the virial radius but which are still associated with a given group. With this physical motivation, we require that for a galaxy to be considered truly isolated, $d_{k}^\mathrm{iso} > 2R_{200,k}$ and $|v_\mathrm{LOS}^\mathrm{iso} - v_\mathrm{LOS}^{\mathrm{group \ cen}, k}|> 3\sigma_{v,k}$ for all $k$ groups containing more than one galaxy. If these criteria are not met, the algorithm attempts to reassign the isolated galaxy as a satellite of the closest group in terms of both relative distance and velocity. A group is considered a candidate host for the isolated galaxy if the following criteria for the distance and velocity of the isolated galaxy relative to the group's central are satisfied:
\begin{eqnarray}
    d^\mathrm{iso} \leq 2R_{200} \\
    |v_\mathrm{LOS}^\mathrm{iso} - v_\mathrm{LOS}^\mathrm{group \ cen}| \leq 3 \sigma_v \\
    M_\ast^\mathrm{iso} < M_\ast^\mathrm{group \ cen}
\end{eqnarray}
If no group meets these criteria, then the isolated galaxy remains isolated. If more than one group passes this selection, then the isolated galaxy is reassigned as a satellite of the group for which the quantity $\sqrt{(d/2R_{200})^2 + (\Delta v / 3\sigma_v)^2}$ is minimal.

\begin{figure*}[h]
    \centering
    \includegraphics[width=\linewidth]{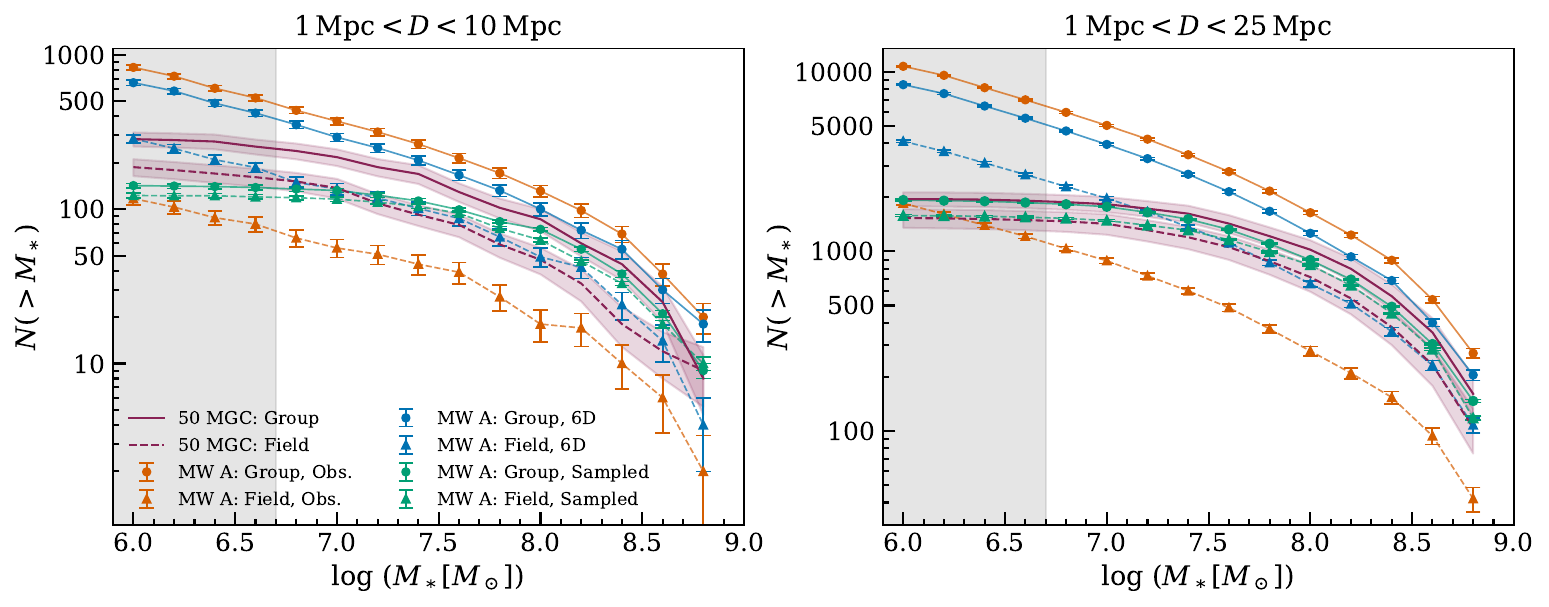}
    \caption{Comparison of cumulative galaxy counts separately for field and group galaxies within (\textit{left}) $1 < D/\mathrm{Mpc} < 10$ and (\textit{right}) $1 < D/\mathrm{Mpc} <25$ when different group finder approaches are implemented. The orange points illustrate the simple `observationally consistent' abundance matching scheme outlined in \ref{sec:A1_obs} when applied to the TNG50 MW A sample. The blue points represent the result of classifying the TNG50 MW A simulated sample based on the 6D abundance matching scheme described in \ref{sec:A1_sim}. The green points are the cumulative galaxy counts for field and group galaxies when the TNG50 MW A data set is sampled $N = 1,000$ times according to the mass completeness of 50 MGC, as described in \ref{sec:groupfinder_masscompleteness}. The completeness is estimated as the fraction $N_\mathrm{obs}/N_\mathrm{sim}$ relative to MW A in mass bins of $0.2$ dex within 25 Mpc. Galaxies are classified according to the group finder described in \ref{sec:A1_obs} for this subsample of the MW A data. Group and field galaxies in the 50 MGC sample (magenta lines) are classified according to the method described in \ref{sec:A1_obs}.}
    \label{fig:SMF-env-comparisons}
\end{figure*}

When classifying groups in the 50 MGC sample, we use the approach outlined here for simulations, with the only difference being that we start with the observables of RA, Dec, velocity, and radial or line-of-sight distance (which may or may not depend on redshift/velocity). The results of applying this method to the TNG50 MW A sample and to the 50 MGC data are shown in Fig. \ref{fig:SMF-env-comparisons}, where we compare the cumulative number counts as a function of mass separated by their environment classification.

\subsubsection{6D Implementation}\label{sec:A1_sim}
In the full 6D implementation, the logic remains the same, but distances are computed from the full 3D position vectors in our galactocentric coordinates, and the velocity criteria becomes a requirement on the 3D peculiar velocities of satellites relative to the central. Mathematically, satellite classification in parts 1 and 2 must satisfy:
\begin{equation}
    \|\mathbf{D}_\mathrm{sat} - \mathbf{D}_\mathrm{cen}\| \leq R_{200} \ \mathrm{and} \ \|\mathbf{v}_\mathrm{pec}^\mathrm{sat}-\mathbf{v}_\mathrm{pec}^\mathrm{cen}\| \leq 3 \sigma_v
\end{equation}
To be considered an isolated central, a galaxy must satisfy the following criteria with respect to all $k$ group centrals:
\begin{equation}
    \|\mathbf{D}_\mathrm{iso} - \mathbf{D}_{\mathrm{group \ cen},k}\| > 2R_{200,k} \ \mathrm{and} \ \|\mathbf{v}_\mathrm{pec}^\mathrm{iso} - \mathbf{v}_\mathrm{pec}^{\mathrm{group \ cen},k}\| > 3 \sigma_{v,k}\ 
\end{equation}
or it may be reassigned to a group for which:
\begin{eqnarray}
    \|\mathbf{D}_\mathrm{iso} - \mathbf{D}_\mathrm{group \ cen}\| \leq 2R_{200} \\
    \|\mathbf{v}_\mathrm{pec}^\mathrm{iso} - \mathbf{v}_\mathrm{pec}^\mathrm{group \ cen}\| \leq 3 \sigma_v \\
    M_\ast^\mathrm{iso} < M_\ast^\mathrm{group \ cen}
\end{eqnarray}
If more than one group passes this selection, the isolated galaxy is assigned as a satellite to the group for which the quantity $\sqrt{(\|\mathbf{D}_\mathrm{iso} - \mathbf{D}_\mathrm{group \ cen}\|/2R_{200})^2 + (\|\mathbf{v}_\mathrm{pec}^\mathrm{iso} - \mathbf{v}_\mathrm{pec}^\mathrm{group \ cen}\| / 3\sigma_v)^2}$ is minimal.

When all position information is available, we can additionally reduce the computational complexity of this algorithm; in each of the three parts, a k-d tree is built once from the positions of all candidate subhalos using the \verb|Aboria| C++ library. During each iteration, the tree is searched for all nearby candidates within a search radius large enough to ensure that classification is insensitive to this chosen radius (at least $3R_{200}$). 

The resulting number counts by group and field classification when the 6D group finder is applied to the TNG50 MW A sample are shown in Fig. \ref{fig:SMF-env-comparisons}. In the 6D implementation, fewer than 40 satellites are reclassified in each MW sample in part 2 of the algorithm, while between $400-600$ satellites are reclassified in the observationally consistent implementation across the three MW samples. In part 3, between $3,000-4,000$ isolated galaxies are reclassified as satellites in the 6D implementation, compared to $4,000-5,000$ in our observationally consistent approach. The higher number of reclassified galaxies in the observationally consistent approach is expected because we base classification on the line-of-sight velocity component, instead of the full 3D velocity as in the 6D case. This leads to the well-known effect of redshift-space distortion, introducing greater uncertainty in group classification.

\begin{figure}[h]
    \centering
    \includegraphics[width=\linewidth]{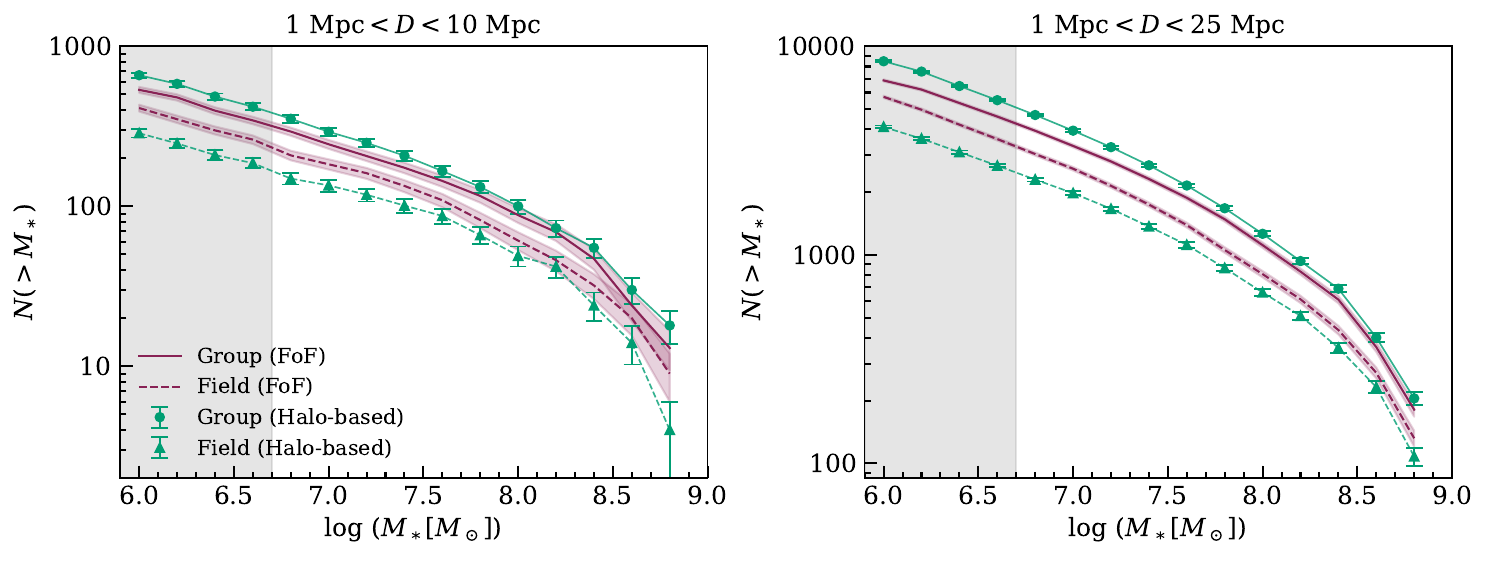}
    \caption{Comparison of the 6D group classification method outlined in \ref{sec:A1_sim} to the results of the FoF algorithm for the MW A simulated galaxy sample, within (\textit{left}) $1 \leq D/\mathrm{Mpc} \leq 10$ and (\textit{right}) $1 \leq D/\mathrm{Mpc} \leq 25$. Field galaxies within the FoF classification are defined as those which are the single galaxy in a FoF group after excluding all subhalos that do not meet our mass criteria of $\geq 25$ stellar particles, or are not of cosmological origin. Group galaxies within the FoF classification are all galaxies belonging to FoF groups with more than one subhalo meeting our criteria.}
    \label{fig:FoF-comparison}
\end{figure}

This 6D classification produces fewer isolated galaxies than is obtained from the FoF algorithm (Fig. \ref{fig:FoF-comparison}). In both volumes, there are $\sim 40\%$ more isolated galaxies with masses $10^6 \leq M_\ast/M_\odot \leq 10^9$ in the FoF classification compared to our simple halo-based 6D classification. The difference is almost certainly a consequence of our stricter requirement on isolated galaxies.

\subsubsection{Sensitivity to Mass Completeness}\label{sec:groupfinder_masscompleteness}

The results of the observationally consistent classifier and the 6D classifier (Fig. \ref{fig:SMF-env-comparisons}) are notably different. The much lower number of isolated galaxies predicted by the observationally consistent approach compared to the 6D approach suggests that algorithmic choices may be deflating the number of isolated galaxies. If we accept the 6D classification of the simulation data as a ground truth for the local environment in 10 and 25 Mpc, then the number counts associated with this group finder in Fig. \ref{fig:SMF-env-comparisons} suggest that we should find similar numbers of isolated galaxies and galaxies in groups in both the 50 MGC and TNG50 samples within the regimes where the 50 MGC is mass-complete. This motivated us to explore the sensitivity of our observationally consistent approach to survey completeness. 

The 50 MGC data is largely complete with respect to the MW A sample in the regimes of $M_\ast \gtrsim 10^{8.2} M_\odot$ for $1 < D/\mathrm{Mpc} < 25$ and $M_\ast \gtrsim 10^{7.2} M_\odot$ for $1 < D/\mathrm{Mpc} < 10$ (Fig. \ref{fig:completeness_Frac}). Thus, for a perfect group finder which is insensitive to mass completeness, group and isolated galaxy counts in the simulations would be unchanged in these regimes, up to statistical variance, when the data is down-sampled according to observational mass completeness. Here we assume that the simulated galaxy samples form ground-truth predictions for the observed nearby galaxy population, so that the relative proportion of galaxies in groups and in the field would be identical between the simulated and observed samples in the mass-complete regime in the ideal case.

We estimate the completeness of the 50 MGC as a function of mass according to $f_c(M_\ast) \equiv N_\mathrm{obs}(M_\ast)/N_\mathrm{sim}(M_\ast)$, evaluated within mass bins of $0.2 \ \mathrm{dex}$ and within $D < 25 \ \mathrm{Mpc}$. Within $1 \ \mathrm{Mpc}$, we take $f_c = 1$ for all masses. We evaluate $f_c$ with respect to the entire MW A sample as was done in Fig. \ref{fig:completeness_Frac}, without imposing any assumption about the mass completeness as a function of group or field classification. We then randomly sample all masses $M_\ast > 10^6 \ M_\odot$ in the TNG50 MW A sample according to $f_c$. If $f_c(M_\ast) > 1$ for any mass bin, we include all TNG50 galaxies within that mass range. Within $1 < D/\mathrm{Mpc} < 25$, the 50 MGC sample is actually overabundant with respect to the MW A sample between $10^{9} \lesssim M_\ast/M_\odot \lesssim 10^{11}$ (Fig. \ref{fig:completeness_Frac}), so the resulting MW A sample is actually smaller than the 50 MGC sample overall. The group finder of Appendix \ref{sec:A1_obs} was then applied to $N = 1,000$ random samples.

The results of the sampling on the group and field counts as a function of mass are shown in Fig. \ref{fig:SMF-env-comparisons}. We show the median and median absolute deviation (MAD) of the distribution of number counts per mass bin after all group finder runs. The obvious changes in cumulative number counts between these results and the observationally consistent group finder results (Fig. \ref{fig:SMF-env-comparisons}) in the mass-complete regimes ($M_\ast \gtrsim 10^{7.2} \ M_\odot$ for $D < 10$ Mpc, $M_\ast \gtrsim 10^{8.2} \ M_\odot$ for $D < 25$ Mpc) suggest that the approach outlined in Appendix \ref{sec:A1_obs} is highly sensitive to mass completeness. 

Although numerous studies have used mock catalogs from simulations to quantify biases in recovering group properties and membership for observational group finders (e.g. \citealt{Berlind_2006, Robotham2011, Duarte2014, Campbell_2015, CasoVega-Martinez_2019, Marini_2025}), limited work has been done to assess the mass dependence of observational group finders using the environments of LG or MW analogues in cosmological hydrodynamical simulations. Because we have full 6D phase space information for galaxies down to $M_\ast \sim 10^6 \, M_\odot$ within 25 Mpc of MW analogues in cosmological simulations, we are uniquely equipped to evaluate the mass dependence of group finders in the regimes where corrections for survey completeness are not straightforward to apply. In the next section, we consider an adaptation of the \citet{Yang_groupfinder_2005} approach for group classification to minimize the mass bias inherent to observational group finders. 

\subsection{Density-Contrast-Based Classification}\label{sec:groupfinder_method}

In this section, we present our procedure for identifying galaxy groups in both simulations and observations, which is based on the method of \citet{Yang_groupfinder_2005}. The group finding algorithm of \citet{Yang_groupfinder_2005} classifies galaxy groups based on their number density contrast in redshift space. This is attractive for our purposes of developing a group finding algorithm which is less sensitive to catalog incompleteness because, to first order, the typical number density contrast of galaxy groups should roughly scale with the number density of galaxies in the catalog. Henceforth, we adapt the \citet{Yang_groupfinder_2005} algorithm.

When applied to flux-limited surveys, it is common for group finding algorithms to take into account the incompleteness as a function of luminosity or redshift when computing group membership (e.g. \citealt{Yang_groupfinder_2005, Knobel2012, Lim2017, Tinker2021}). This is particularly relevant when assigning halo masses, in which the mass of a preliminarily assigned group may be underestimated unless some care is taken in accounting for this incompleteness.  After preliminary assignment of potential groups using an FoF algorithm, \citet{Yang_groupfinder_2005} weight the individual luminosities of galaxies in each group by the luminosity-dependent incompleteness, and then assign a halo mass from the weighted total. \citet{Tinker2021} account for luminosity-dependent incompleteness via an effective volume correction, which is used to estimate the cumulative number density of halos at a particular redshift given an assumed halo mass function. When analyzing a catalog like 50 MGC, however, it is not straightforward to derive a single flux limit for the whole sample because it consists of multiple different surveys. Since stellar masses are assigned in a self-consistent manner in the \citet{Ohlson_2024} catalog, we can use this as our observable and hence as our tracer of catalog completeness.

In our adaptation of the \citet{Yang_groupfinder_2005} approach, we systematically account for observational incompleteness by scaling the threshold number density contrast required for assignment of satellites to a given group. We assume that dark matter halos follow a universal NFW density profile \citep{NFW1997}, and that the phase space density of galaxies follows that of the underlying dark matter distribution. The NFW density profile is parameterized by a scale radius, $r_s$, and a halo concentration parameter, $c_{200}$, which we define in terms of $R_{200}$ so that $c_\mathrm{200} = R_{200}/r_s$. Similar to \citet{Yang_groupfinder_2005}, we adopt a concentration-mass relation so that we can determine $c_{200}$ from $M_h$, and hence determine the free parameter $r_s$. We choose the recent relation of \citet{DiemerJoyce2019}, adapted to our cosmology.

The 2D projected surface density of an NFW profile at $z=0$, as expressed by \citet{Yang_groupfinder_2005} and \citet{Tinker2022_groupfinder}, is given by:
\begin{equation}
    \Sigma(d) = 2 \ r_s \ \overline\delta \ \rho_{c,0}  \ f(d/r_s)
\end{equation}
where
\begin{eqnarray}
    f(x) = 
    \begin{cases}
    \frac{1}{x^2-1}\left(1 - \frac{\ln [(1+\sqrt{1-x^2})/x]}{\sqrt{1-x^2}} \right) \hspace{5pt} \mathrm{if} \ x<1 \\
    \frac{1}{3} \hspace{115pt} \mathrm{if} \ x=1 \\
    \frac{1}{x^2-1}\left(1 - \frac{\mathrm{arctan} \sqrt{x^2 - 1} }{\sqrt{x^2 - 1}}\right) \hspace{18pt} \mathrm{if} \ x>1
    \end{cases}
\end{eqnarray}
and 
\begin{equation}
    \overline\delta = \frac{200}{3} \frac{c_{200}^3}{\ln (1+c_{200}) - c_{200}/(1+c_{200})}
\end{equation}
The density contrast in velocity space between a given galaxy and the group central, which we still assume resides at the center of the halo, is given by:
\begin{equation}
    P(d, \Delta v) = H_0 \frac{\Sigma(d)}{\rho_{c,0}} p(\Delta v)
\end{equation}
where the relative line-of-sight velocity $\Delta v = v_\mathrm{LOS}^\mathrm{sat} - v_\mathrm{LOS}^\mathrm{cen}$ is assumed to follow a Gaussian distribution with variance $\sigma_v^2 \equiv V_\mathrm{vir}^2/2$:
\begin{equation}
    p(\Delta v) = \frac{1}{\sqrt{2\pi}\sigma_v} \exp \left[\frac{-(\Delta v) ^2}{2\sigma_v^2}\right]
\end{equation}

The key assumption is that the number density of galaxies traces the mass density of dark matter, so that $P(d, \Delta v)$ may be interpreted as a galaxy number density contrast in velocity space. With this in mind, we choose a physically-motivated value $B$ to serve as the threshold above which a galaxy may be assigned to a halo. \citet{Yang_groupfinder_2005} motivate their choice of this threshold ``background" term by considering the approximate redshift-space density contrast at the edge of a halo. We similarly choose:
\begin{equation}\label{eq:B_definition}
    B_0 \equiv \frac{\rho_\mathrm{vel}(R_{200})}{\rho_{c,0}} = \frac{\rho(R_{200})}{\rho_{c,0}} \frac{(4\pi/3)R_{200}^3}{\pi R_{200}^2 \sigma_v/H_0} = \frac{4 \sqrt{2}}{3}\frac{\rho(R_{200})}{\rho_{c,0}} \frac{R_{200}}{V_\mathrm{vir}}
\end{equation}
where $\rho(R_{200})$ is computed from the NFW density profile \citep{NFW1997}. $B_0$ is implicitly dependent on mass via the halo concentration parameter. Its explicit dependence on $R_{200}/V_\mathrm{vir}$ allows for a straightforward scaling of the threshold for group inclusion. For example, an isolated galaxy will be reassigned as a satellite if $P(d, \Delta v) < (2/3)B_0$ relative to at least one multi-member group, where the velocity-space edge of the halo is set at $2R_\mathrm{200}$ and $3 \sigma_v$ as in Appendix \ref{sec:A1_obs}.

\citet{Yang_groupfinder_2005} calibrate $B$ based on mock samples such that contamination is minimized and completeness maximized, yielding $B = 10$. \citet{Tinker2022_groupfinder} assumes the same value. We find the differences between using $B = 10$ and the mass-dependent $B_0$ above are minor, with the constant $B$ approach producing slightly more isolated galaxies; $40 \%$ of galaxies in 25 Mpc of MW A are classified as isolated when a constant $B$ is used, compared to $37 \%$ with the mass-dependent $B_0$. The use of $B_0$ produces results that are closer to the 6D implementation, in which $34\%$ of galaxies are marked as isolated in 25 Mpc of MW A, and so we retain our mass-dependent definition of $B$ in Eq. \ref{eq:B_definition}. 

When we apply our algorithm to observations, we scale $B$ such that:
\begin{equation}
    B = f_C B_0
\end{equation}
where we have defined the total completeness fraction ($f_C$):
\begin{equation}\label{eq:B_scaling}
    f_C \equiv \begin{cases}
    \frac{n_\mathrm{obs}}{n_\mathrm{sim}} = \frac{N_\mathrm{obs}}{N_\mathrm{sim}}\frac{V_\mathrm{sim}}{V_\mathrm{obs}} \hspace{5pt} \mathrm{for \ observations} \\
    1 \hspace{68pt} \mathrm{for \  simulations}
    \end{cases}
\end{equation}
The quantities $n_\mathrm{obs}$ and $n_\mathrm{sim}$ are the mean galaxy number densities within the whole observational and simulation samples, respectively, that are considered by the group finder. Relative to each of the three MW samples, $f_C = n_\mathrm{obs}/n_\mathrm{sim} \approx 1/9$. Physically, $B$ is interpreted as a background number density contrast, analogous to a linking length in FoF algorithms, so to first order, $B$ should scale with the available number of tracers of the underlying dark matter density field.

The approach presented in Appendix \ref{sec:A1_obs} is modified in the following manner for our density-contrast-based algorithm:

\textbf{Part 1}. A satellite is classified as part of a group if, relative to the group center:
\begin{equation}
    P(d, \Delta v) \geq  \frac{1}{3} f_C B_0
\end{equation}
where $f_C$ is defined separately for simulations and observations as in Eq. \ref{eq:B_scaling}.

\textbf{Part 2}. Satellites may be reclassified to any group $k$ for which $P_k(d, \Delta v) \geq (1/3) f_C B_{0,k}$. Of this subset, the satellite is classified to the group for which $P$ is maximal.

\textbf{Part 3}. Isolated galaxies must satisfy the criteria that $P_k(d, \Delta v) < (2/3) f_C B_{0,k}$ relative to all $k$ multi-member groups. If this is not true, then there is at least one multi-member group for which $P \geq (2/3) f_C B_0$. If more than one group passes this criterion, then the isolated galaxy is assigned to the group for which $P$ is maximal.

The resulting group and field galaxy counts for the MW A sample are presented in the main text in Fig. \ref{fig:SMF-env}. Our density-contrast-based approach classifies a similar amount of group and field galaxies as the 6D group finder (Fig. \ref{fig:SMF-env-comparisons}), validating the robustness of this approach. 

\section{Milky Way Analogues B and C}\label{sec:A2_MWBC}

In this appendix, we include analyses of the MW B and C sample properties that were omitted from the main text for clarity of presentation. In \S \ref{results:counts} we demonstrated that the difference in galaxy counts among the three MW samples within 10 Mpc could be explained by their differing local dark matter densities, as traced by the number of bright galaxies $N(> 10^{9.5} M_\odot)$ in each volume (Fig. \ref{fig:DMdensity}). For distances out to 25 Mpc, the number counts converged. The greatest differences in sample statistics for the three MW realizations in 10 Mpc can be reduced to a scaling factor, so we selected MW A as the reference sample. This statement is evident in the four figures presented below.

In Fig. \ref{fig:completeness_BC}, we show the calculated fraction of the LVDB and 50 MGC samples relative to the MW B and C samples, which we define as the completeness fraction of these surveys relative to the simulated galaxy samples. While there are differences within $1$ Mpc and $10$ Mpc, the estimated mass completeness of the 50 MGC with respect to the three MW samples converges within $D < 25~\mathrm{Mpc}$. In Fig. \ref{fig:env-all}, we present galaxy counts by group vs. field for all three MW samples, and in Fig. \ref{fig:color-all}, we show number counts for red/quenched and blue/star-forming galaxies in the three samples. Finally, in Fig. \ref{fig:quenched_fraction-all}, we demonstrate broad consistency between the quenched fractions for all three MW samples. As before, the differences within 10 Mpc that are illustrated in these figures can be attributed to cosmic variance and different local dark matter densities. In Fig. \ref{fig:quenched_fraction-all}c and \ref{fig:quenched_fraction-all}d, the group quenched fraction is generally higher for MW analogues that have higher mass densities within $1 < D/\mathrm{Mpc} < 10$ (Table \ref{tab:TNG50_MW_properties}), which is an expected consequence of the fact that environmental quenching mechanisms are thought to be more effective in denser large-scale environments.

\begin{figure}[h]
    \centering
    \includegraphics[width=\linewidth]{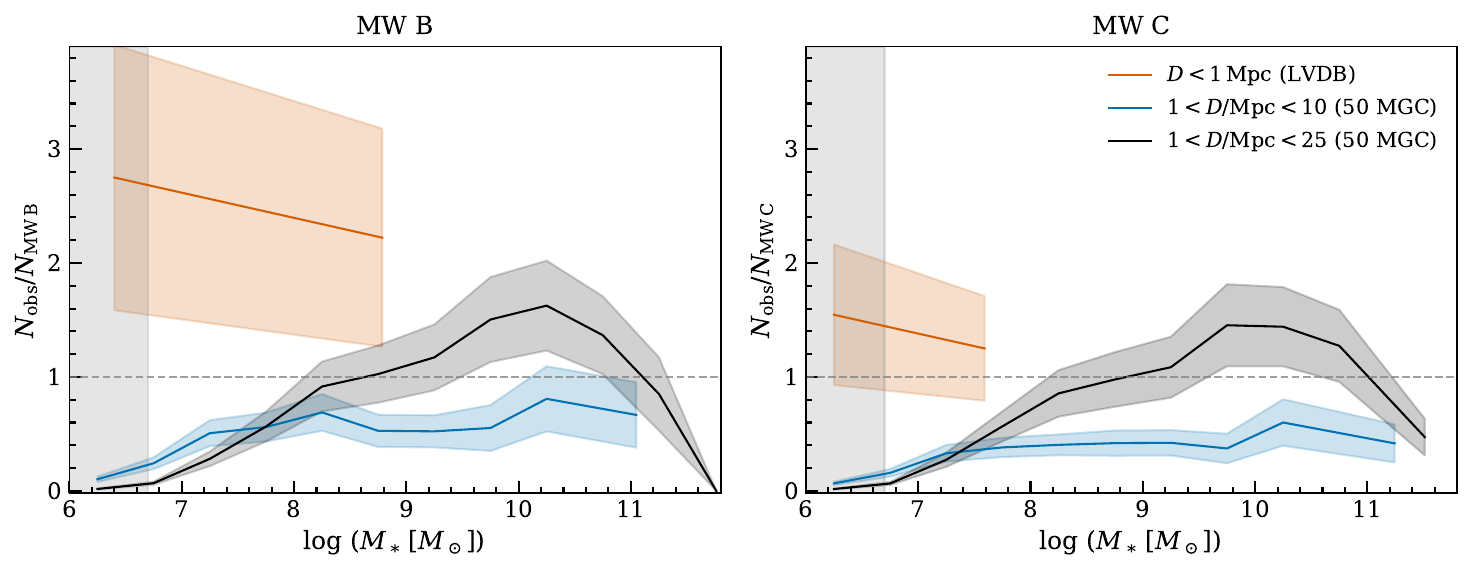}
    \caption{Ratio of galaxy counts within the LVDB and 50 MGC samples relative to the (\textit{left}) TNG50 MW B and (\textit{right}) MW C samples as a function of mass and distance (similar to Fig. \ref{fig:completeness_Frac}). Adaptive mass binning is applied to ensure $N_\mathrm{sim} \geq 10$ as in Fig. \ref{fig:completeness_Frac}, although this requirement is relaxed to $N_\mathrm{sim} \geq 8$ for the relatively underdense MW B environment within $1~\mathrm{Mpc}$ at these masses. Errors are estimated from the tabulated measurement uncertainties for the observational data and the Poisson error of both the simulation and observational data.}
    \label{fig:completeness_BC}
\end{figure} 

\begin{figure}[hb!]
    \centering
    \includegraphics[width=\linewidth]{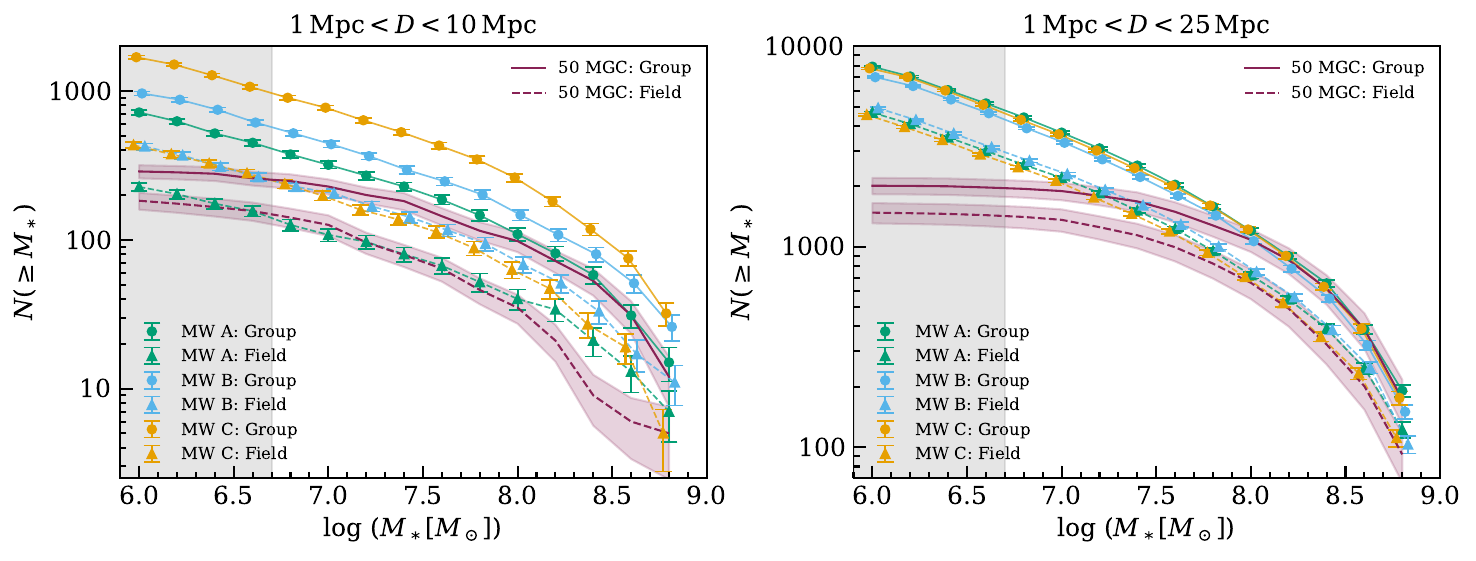}
    \caption{Similar to Fig. \ref{fig:SMF-env}, but including cumulative number counts subdivided by group/field classification for MW B and C within ($\textit{left}$) 10 Mpc and ($\textit{right}$) 25 Mpc. A small offset in $\log M_\ast$ is applied to all points relative to MW A for easier visualization.\\}
    \label{fig:env-all}
\end{figure}

\begin{figure}[hb!]
    \centering
    \includegraphics[width=\linewidth]{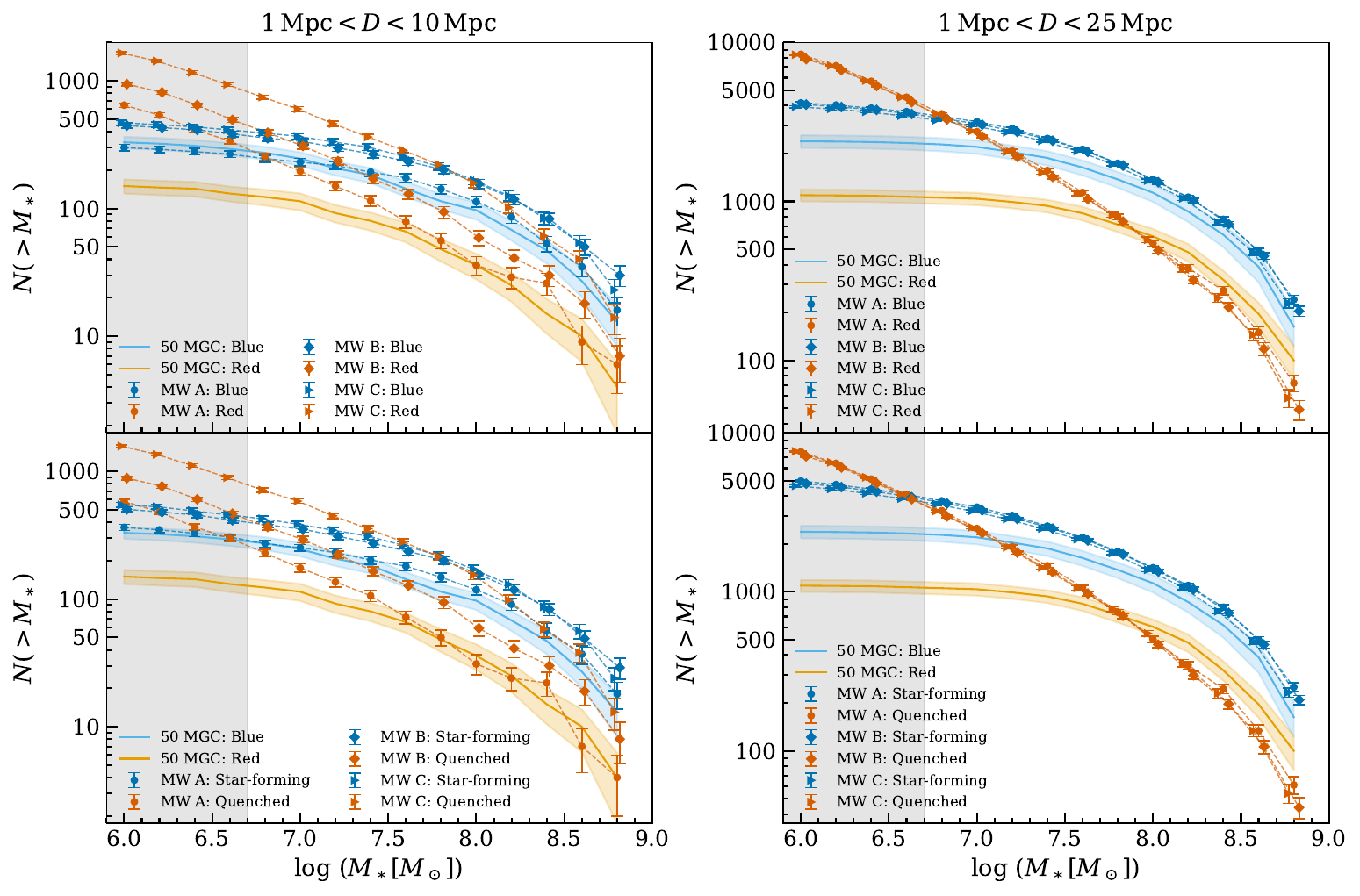}
    \caption{Similar to Fig. \ref{fig:SMF-color}, but including number counts for the TNG50 MW B and C environments for (\textit{top row}) blue and red populations and (\textit{bottom row}) star-forming and quenched populations. A small horizontal offset is applied to the MW B and C points relative to MW A for easier visualization.\\}
    \label{fig:color-all}
\end{figure}

\begin{figure}[h]
    \centering
    \includegraphics[width=\linewidth]{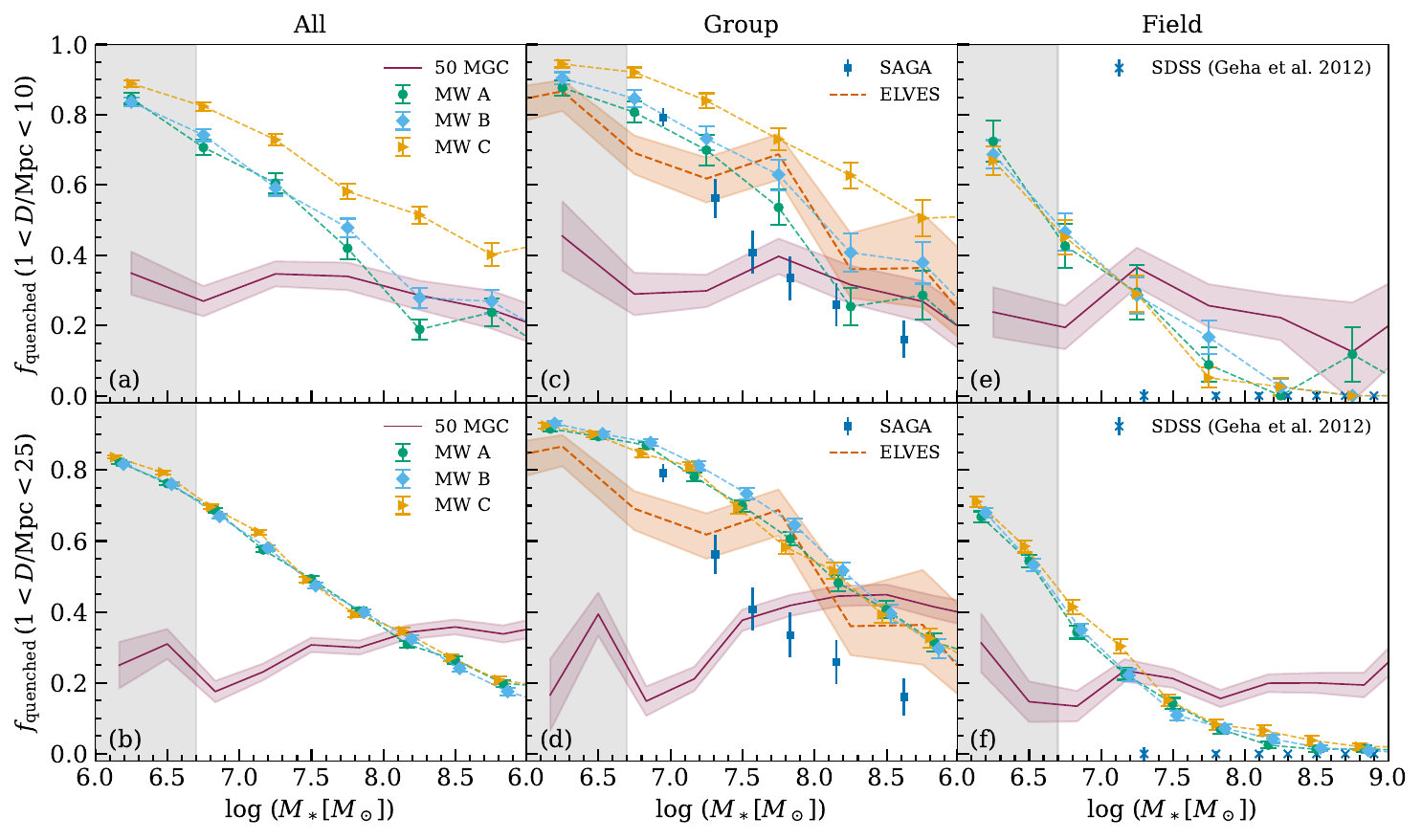}
    \caption{Similar to Fig. \ref{fig:quenched_fraction}, but including the quenched fractions as a function of mass for the MW B and C samples. In the bottom row, a small horizontal offset is applied to the MW B and C points relative to MW A for easier visualization.}
    \label{fig:quenched_fraction-all}
\end{figure}

\section{Mass Proxy Comparison}\label{sec:AppC}

\begin{figure}
    \centering
    \includegraphics[width=0.5\linewidth]{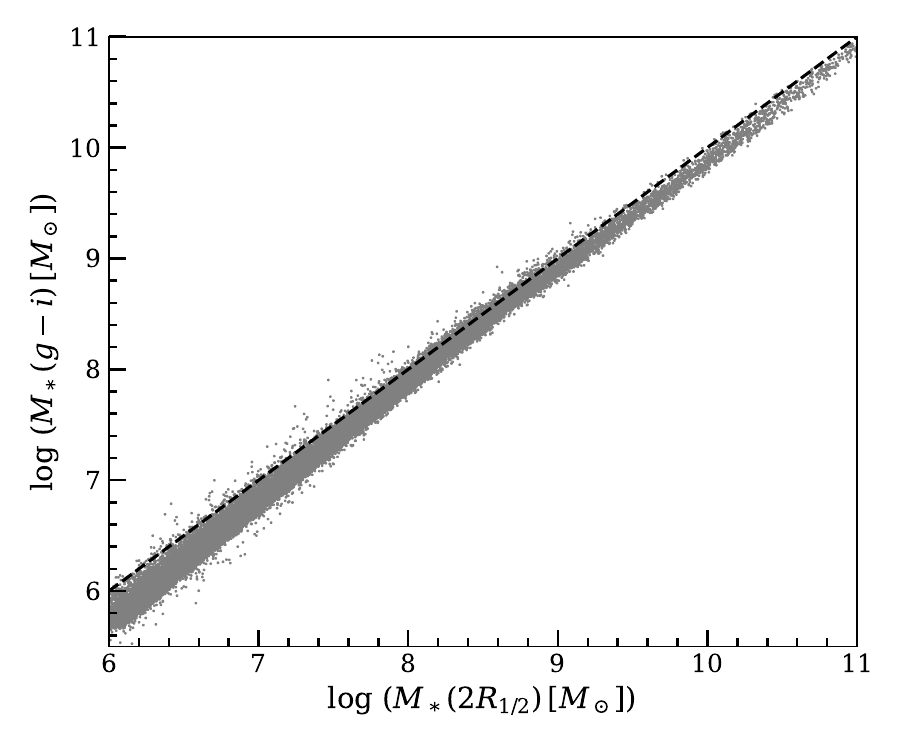}
    \caption{Our definition of the TNG50 masses as the total stellar mass within $2R_\mathrm{1/2}$ is compared to the $g-i$-derived stellar mass definition considered in comparison to 50 MGC. We apply the relation of \citet{Taylor2011_GAMA}, using TNG photometry. The 1:1 line is shown for visualization (dashed black line), and only subhalos with $\geq$ 25 stellar particles are shown (gray points).}
    \label{fig:mass_rhalf_vs_g-i}
\end{figure}

In Fig. \ref{fig:DMdensity}, we used TNG photometry to define the stellar masses for the TNG50 subhalos according to the $(g-i)-M_\ast/L_i$ relation of \citet{Taylor2011_GAMA} for a consistent comparison with the 50 MGC data. In Fig. \ref{fig:mass_rhalf_vs_g-i}, we show that the stellar masses, when defined in this way, generally agree with the mass definition adopted throughout the remainder of the paper which is based on the total stellar mass within $2R_\mathrm{1/2}$. The color-derived masses skew slightly lower, with an average relative offset of $-0.18~\mathrm{dex}$.

\end{document}